\documentclass[11pt,usletter]{article}
\usepackage{jheppub}
\usepackage{verbatim}
\usepackage{graphicx}
\usepackage{hyperref}
\usepackage{color}
\usepackage{amsmath}
\usepackage{tikz}
\usepackage{xcolor}

\newcommand{\M}{\mathcal{M}}

\newcommand{\mO}{\mathcal{O}}

\newcommand{\tx}{\text{x}}
\newcommand{\ty}{\text{y}}

\newcommand{\zb}{\bar z}
\newcommand{\bL}{{\bf L}}

\newcommand{\mat}[4]{\left(\begin{array}[c]{cc}
#1 & #2\\
#3 & #4
\end{array}\right)}

\usepackage[OT2,T1]{fontenc}
\DeclareSymbolFont{cyrletters}{OT2}{wncyr}{m}{n}

\title{2D Ising Field Theory in a Magnetic Field:\\
The Yang-Lee Singularity}
\author{Hao-Lan Xu,}
\author{and Alexander Zamolodchikov}

\affiliation{C.N. Yang Institute for Theoretical Physics, State University of New York, Stony Brook, NY 11794-3840, USA}

\emailAdd{hao-lan.xu@stonybrook.edu}
\emailAdd{alexander.zamolodchikov@stonybrook.edu}
\begin{document}

\tikzset{every picture/.style={line width=0.75pt}} %set default line width to 0.75pt
\begin{flushright}
YITP-SB-2022-12\\
\end{flushright}
\abstract{
We study Ising Field Theory (the scaling limit of Ising model near the Curie critical point) in pure imaginary external
magnetic field. We put particular emphasis on the detailed structure of the Yang-Lee edge singularity. While the leading singular behavior is controlled by the Yang-Lee fixed point ($=$ minimal CFT ${\cal M}_{2/5}$), the fine structure of the subleading singular terms is determined by the effective action which involves a tower of irrelevant operators. We use numerical data obtained through the "Truncated Free Fermion Space Approach" to estimate the couplings associated with two least irrelevant
operators. One is the operator $T{\bar T}$, and we use the universal properties of the $T{\bar T}$ deformation to fix the contributions of higher orders in the corresponding coupling parameter $\alpha$. Another irrelevant operator we deal with is the descendant $L_{-4}{\bar L}_{-4}\phi$ of the relevant primary $\phi$ in ${\cal M}_{2/5}$. The significance of this operator is that it is the lowest dimension operator which breaks integrability of the effective theory. We also establish analytic
properties of the particle mass $M$ ($=$ inverse correlation length) as the function of complex magnetic field.
}

\maketitle
\flushbottom

\section{Introduction}

The scaling behavior of the 2D Ising model in the external magnetic field $H$ near its ferromagnetic critical point $(T,H)=(T_c,0)$ is of much interest since it represents the basic universality class which includes, in particular, the Curie criticality in the axial ferromagnet, as well as the liquid-vapor critical point of simple gasses in two dimensions \cite{mccoy2013two}. It is also of much interest as the model of quantum field theory as it exhibits a range of interesting phenomena \cite{mccoy2013two,mccoy1978two,fonseca2003ising}. At zero magnetic field the model admits, of course, an exact solution because in this case it is reduced to the theory of free Majorana fermions in 2D Euclidean space-time. At generic nonzero magnetic field the model is not free, and generally not integrable.
This work is a continuation of an extended project of studying the analytic properties of the theory (i.e. its
thermodynamic and correlation characteristics) at complex values of the parameters, initiated in Ref.\cite{fonseca2003ising}.

\subsection*{Ising Field Theory}

The (Euclidean) quantum field theory which appears in this scaling limit is generally referred to as the Ising Field Theory (IFT) \cite{Wu:1975mw}. It can be alternatively defined as the Renormalization Group (RG) flow out of the Ising fixed point (described by the minimal CFT $\mathcal{M}_{3/4}$ \cite{belavin1984infinite}) generated by its two relevant scalar operators -  the "energy density" $\varepsilon(x)$ and the "spin density" $\sigma(x)$. This definition can be expressed via the formal action
\begin{eqnarray}\label{ift}
\mathcal{A}_\text{IFT} = \mathcal{A}_\text{ICFT} + \frac{m}{2\pi}\int\,\varepsilon(x)\,d^2 x
+ h\,\int\,\sigma(x)\,d^2 x
\end{eqnarray}
where $\mathcal{A}_\text{ICFT}$ stands for the formal action of the Ising fixed point theory - the minimal CFT $\mathcal{M}_{3/4}$ with the Virasoro central charge $c_\text{Ising}=\frac{1}{2}$. The coupling parameters are related to the deviations from the critical point in the scaling limit, $m \sim T_c-T$, $h \sim H$. Their exact normalizations depend on the normalizations of the fields $\varepsilon(x)$ and $\sigma(x)$; we fix the latter by the short-distance asymptotic behavior of the two-point correlation functions
\begin{eqnarray}
\langle\,\varepsilon(x)\varepsilon(x')\,\rangle \to |\,x-x'|^{-2}\,, \qquad
\langle\,\sigma(x)\sigma(x')\,\rangle \to |\,x-x'|^{-1/4}
\end{eqnarray}
as $|x-x'|\to 0$. Then the parameters $m$ and $h$ have the mass dimensions $m \sim [\text{mass}]$, $h \sim [\text{mass}]^{15/8}$.
Therefore, up to overall scale, the theory depends on a single scaling parameter
\begin{eqnarray}\label{xidef}
\xi \ = \ \frac{h}{|m|^{15/8}}\ \sim \ \frac{H}{|T_c-T|^{15/8}}\,,
\end{eqnarray}
where the last form is to remind the relation to the parameters of the microscopic Ising model. Thus, various thermodynamic and correlation functions depend, apart from the overall scale, on the dimensionless parameter $\xi$. This work is a follow-up to Ref.\cite{fonseca2003ising}, where the analytic properties of these functions at complex values of the scaling parameter\footnote{At real $\xi$ the scaling parameter \eqref{xidef} relates to $\eta=m/h^{8/15}$ defined in \cite{fonseca2003ising} as $\xi=(-\eta)^{-8/15}$, with $\eta$ taking real negative values in the High-T domain.} was considered. Here we concentrate attention on the "High-T" domain $m<0$ ($T>T_c$), where at $h=0$ the $\mathbb Z_2$ symmetry $\sigma \to -\sigma$ is unbroken,
and the thermodynamic characteristics of the theory \eqref{ift} analytically depend on $\xi^2$
\footnote{On the other hand, in the "Low-T domain" $m>0$ the analyticity is broken at $\Re e \xi = 0$, which is the line of the first-order phase transition.}. In what follows we fix the scale by choosing the units in which
\begin{eqnarray}
|m|=1\,;
\end{eqnarray}
in this units $h$ coincides with $\xi$.

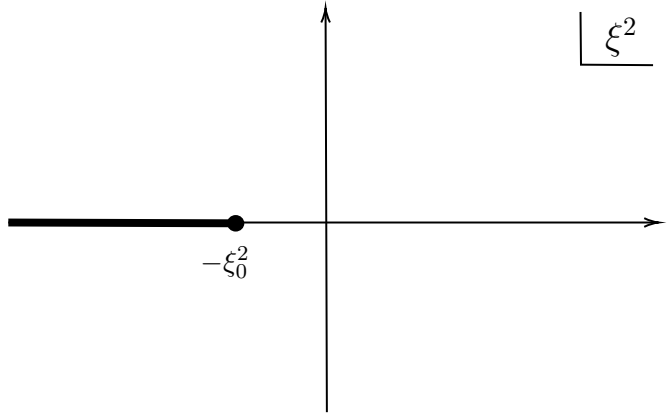
\begin{figure}[ht]
\centering
\begin{tikzpicture}[x=0.5pt,y=0.5pt,yscale=-1,xscale=1]
%uncomment if require: \path (0,896); %set diagram left start at 0, and has height of 896

%Straight Lines [id:da3399239833374874]
\draw    (79.8,671.8) -- (569.8,671.8) ;
\draw [shift={(571.8,671.8)}, rotate = 180] [color={rgb, 255:red, 0; green, 0; blue, 0 }  ][line width=0.75]    (10.93,-3.29) .. controls (6.95,-1.4) and (3.31,-0.3) .. (0,0) .. controls (3.31,0.3) and (6.95,1.4) .. (10.93,3.29)   ;
%Straight Lines [id:da6214634050687411]
\draw    (320.8,815.4) -- (319.81,511.4) ;
\draw [shift={(319.8,509.4)}, rotate = 449.81] [color={rgb, 255:red, 0; green, 0; blue, 0 }  ][line width=0.75]    (10.93,-3.29) .. controls (6.95,-1.4) and (3.31,-0.3) .. (0,0) .. controls (3.31,0.3) and (6.95,1.4) .. (10.93,3.29)   ;
%Straight Lines [id:da7599973469533079]
\draw [line width=3]    (79.8,671.8) -- (251.8,672.4) ;
%Flowchart: Summing Junction [id:dp8744190220896932]
\draw   [fill={rgb, 255:red, 0; green, 0; blue, 0 }  ,fill opacity=1 ] (246.02,672.4) .. controls (246.02,669.36) and (248.61,666.9) .. (251.8,666.9) .. controls (254.99,666.9) and (257.58,669.36) .. (257.58,672.4) .. controls (257.58,675.44) and (254.99,677.9) .. (251.8,677.9) .. controls (248.61,677.9) and (246.02,675.44) .. (246.02,672.4) -- cycle ; \draw   (247.71,668.51) -- (255.89,676.29) ; \draw   (255.89,668.51) -- (247.71,676.29) ;
%Shape: Right Angle [id:dp5958880712746788]
\draw   (567.56,552.67) -- (513.04,552.31) -- (512.67,512.39) ;

% Text Node
\draw (223,687.8) node [anchor=north west][inner sep=0.75pt]    {$-\xi _{0}^{2}$};
% Text Node
\draw (529,514.67) node [anchor=north west][inner sep=0.75pt]  [font=\Large]  {$\xi ^{2}$};
\end{tikzpicture}
\caption{Analyticity structure of $F(\xi^2)$ and $M(\xi^2)$ at complex $\xi^2$ plane in the high temperature domain ($T>T_c$). The branching point located at $\xi^2 = -\xi_0^2$ is the YL singularity. The branch cut extends from $-\xi^2_0$ to $-\infty$, which represents the line of first order phase transition.}\label{complexxiphasediagram}
\end{figure}

The theory \eqref{ift} is massive at all real values of $\xi$. The number of stable particles depends on $\xi$, while their masses change continuously with scaling parameter $\xi$ \cite{mccoy1978two,fonseca2003ising}. In this work we are interested in the mass of the lightest particle, denoted here as $M$, which defines the correlation length $R_c = M^{-1}$. Also, we denote $F$ the bulk vacuum energy density defined, as usual, as the infinite 2D volume limit of $-\log(Z)/V$ \footnote{Under our choice of units $|m|=1$ the function $F(\xi^2)$ simply relates to the scaling function $G(\xi)$ defined in \cite{fonseca2003ising}, $G(\xi)=F(\xi^2)$.}. In statistical mechanics application of the theory \eqref{ift} $F$ is interpreted as the specific free energy. These quantities exhibit analytic dependence on $\xi^2$ for all $\xi^2 \ge 0$, and they can be analytically continued to complex values of $\xi^2$. Thus defined functions $M(\xi^2)$ and $F(\xi^2)$ are analytic on the whole complex $\xi^2$-plane with the branching point at certain negative value $\xi^2=-\xi_0^2 \approx - 0.03583\dots$. This singularity appears as the result of condensation of the Yang-Lee zeros in the thermodynamic limit, and it is known as the Yang-Lee edge singularity. One defines the principal branch by introducing the branch cut along the real axis, from $-\infty$ to $-\xi_0^2$, as shown in Fig.\ref{complexxiphasediagram}. Physically, this branch cut represents the line of the first order phase transitions, whereas the branching point $-\xi_0^2$ is critical (the inverse correlation length $M$ vanishes at this point, see \cite{fisher1978yang} and our discussion below).

%\begin{figure}[ht]
%\centering
%\includegraphics[width=10cm]{xxiplane-1.eps}
%\caption{Phase diagram at complex $\xi$} \label{Fig.1}
%\end{figure}

\subsection*{Yang-Lee Singularity and Yang-Lee QFT}

When $m$ is real and negative (i.e. $T>T_c$) and $h$ is taken pure imaginary, $h=ig$, \eqref{ift} defines a quantum field theory which, albeit being non-unitary, exhibits many reality properties of conventional QFT\footnote{ This is related to the "pseudo-hermiticity" of the theory: At pure imaginary $h$ its Hamiltonian satisfies $H^\dagger = S H S$, where the involution $S,\ S^2=1$, acts by changing the sign of the spin density $\sigma$. The involution $S$ generates an indefinite metric in the space of states. As the result, some quantities which in unitary QFT are interpreted as probabilities (such as cross sections) may take negative values.}. In particular, when  $\xi^2$ is negative but $\xi^2+\xi_0^2 >0$, the theory \eqref{ift} has unique ground state with real energy density $F$, and it gives rise to the particle theory having a single particle of a real mass $M$, with non-trivial scattering theory. The mass $M(\xi^2)$ vanishes at the Yang-Lee point $\xi^2 = - \xi_0^2$, which therefore is critical \cite{fisher1978yang}. The large-scale behavior of this critical theory is controlled by certain conformal field theory - the Minimal Model $\mathcal{M}_{2/5}$ with the central charge $c_{\text{YL}}=-\frac{22}{5}$ \cite{cardy1985conformal}.  (Below we refer to the corresponding RG fixed point as $\mathcal{A}_\text{YLCFT}$, the notation interchangeable with $\mathcal{M}_{2/5}$.) On the other hand, when $\xi^2+\xi_0^2$ is negative, the theory \eqref{ift} has two vacua $| 0_{\pm}\rangle$ with complex vacuum energy densities $F_{\pm}$ which are complex conjugate to each other, $F_{-} = F_{+}^*$. These vacua are "degenerate" in the sense that the real parts of the associated energy densities are equal. Correspondingly, the space of states of the infinite system splits into two sectors, one for each of the vacuum states $| 0_{\pm}\rangle$, and each involving a rich spectrum of complex-mass "particles".

Let us remind here some basics about the Minimal CFT $\mathcal{M}_{2/5}$. This CFT is non-unitary, and has the negative Virasoro central charge $c_{\text{YL}}=-\frac{22}{5}$. There are two primaries, the identity operator $I$ and the scalar primary $\phi$ with the conformal dimensions $(\Delta_\phi, {\bar \Delta}_\phi)=(-\frac{1}{5},-\frac{1}{5})$. Despite
being non-unitary, this CFT has a real structure. One can choose the normalization of $\phi(x)$ so that all the
coefficients in the conformal OPE
\begin{eqnarray}\nonumber
&&\phi(x) \phi(x')= - |x-x'|^{4/5}\,\left[I + \text{descendants}\right]+\qquad\qquad\\ &&\qquad\qquad\qquad\qquad \qquad |x-x'|^{2/5}\,\mathbb{C}_{\phi\phi}^{\phi}\,\left[\phi(x) + \text{descendants}\right]\label{ylope}
\end{eqnarray}
are real, in particular
\begin{eqnarray}\label{Cphi}
\mathbb{C}_{\phi\phi}^{\phi} = \frac{5^{1/4}}{10\pi}\,\frac{\Gamma^2(\frac{1}{5}) \Gamma(\frac{2}{5})}{\Gamma(\frac{4}{5})}\approx 1.91131...\,.
\end{eqnarray}
Let us stress that we use here the normalization of the primary $\phi$ which differs by the factor of $i$ from the one commonly used in the literature  (e.g. \cite{fisher1978yang},\cite{Zamolodchikov:1990bk}),
\begin{eqnarray}
\phi=i\varphi\,.
\end{eqnarray}
This explains the minus sign in the first term in \eqref{ylope}. Although $\varphi$ is directly related to the Lansau-Ginzburg field of \cite{fisher1978yang}, the advantage of our normalization is that it makes the reality property of the OPE algebra \eqref{ylope} explicit.

The CFT $\mathcal{A}_\text{YLCFT}$ has exactly one relevant operator suitable for generating RG flow out of this fixed point, the field $\phi(x)$ itself. This flow is known as the Yang-Lee QFT (see e.g. \cite{fisher1978yang},\cite{cardy1985conformal},\cite{Cardy:1989fw},\cite{Zamolodchikov:1990bk},\cite{zamolodchikov1990thermodynamic}). It is described by the formal action\footnote{ The absence of the factor $i$ in the perturbation term is related, again, to our normalization of the field $\phi$, as in \eqref{ylope}.}
\begin{eqnarray}\label{ylqft}
\mathcal{A}_\text{YL} = \mathcal{A}_\text{YLCFT} + \lambda\,\int \phi(x) d^2 x
\end{eqnarray}
where, as before $\mathcal{A}_\text{YLCFT}$ is the Minimal CFT $\mathcal{M}_{2/5}$. As the primary field $\phi(x)$ has conformal dimensions $(-\frac{1}{5},-\frac{1}{5})$, the coupling constant $\lambda$ carries the mass dimension $[\lambda] = [mass]^{12/5}$. At $\lambda\neq 0$ the QFT is massive and integrable \cite{zamolodchikov1989integrable}. It inherits much of the reality properties from the Yang-Lee CFT, provided the coupling constant $\lambda$ is chosen real and positive.
At the positive values of $\lambda$, the associated factorizable scattering theory was identified in \cite{Cardy:1989fw}; it involves a single kind of neutral particles with the real mass $M_\text{YL}$,
\begin{eqnarray}\label{myl}
{M_\text{YL}} = C_\text{YL}\,\lambda^{5/12}\,,
\end{eqnarray}
where \cite{zamolodchikov1995mass}
\begin{eqnarray}\label{cyl}
C_\text{YL} = \frac{2^\frac{19}{12}\,\sqrt{\pi}}{5^\frac{5}{16}}\,
\frac{\ \ \left[\Gamma(\frac{3}{5})\Gamma(\frac{4}{5})\right]^{\frac{5}{12}}}
{\Gamma(\frac{2}{3})\Gamma(\frac{5}{6})}  = 2.64294463...\,,
\end{eqnarray}
and with the two-particle S-matrix
\begin{eqnarray}\label{syl}
S_\text{YL}(\theta) =\frac{\sinh\theta + i \sin(2\pi/3)}{\sinh\theta - i\sin(2\pi/3)}\,.
\end{eqnarray}
The vacuum energy density $F_\text{YL}$ of YLQFT \eqref{ylqft} is given by \cite{zamolodchikov1990thermodynamic}:
\begin{eqnarray}\label{fyl}
F_\text{YL} = f_\text{YL}\,M_\text{YL}^2\,, \qquad f_\text{YL}=-\frac{\sqrt{3}}{12}\,.
\end{eqnarray}

The theory remains integrable at negative real $\lambda$ (and indeed at complex $\lambda$ as well), although
its physical content in this regime is still poorly understood. At negative $\lambda$ the QFT \eqref{ylqft} has two ground states, $| 0_\pm\,\rangle$ (the phenomenon which can be interpreted as the "spontaneous breakdown" of the symmetry $\phi \to \phi^*$), the associated vacuum energy densities $F_{\pm}$ being complex conjugate to each other.

\subsection*{Renormalization Group Flow and Effective Action}

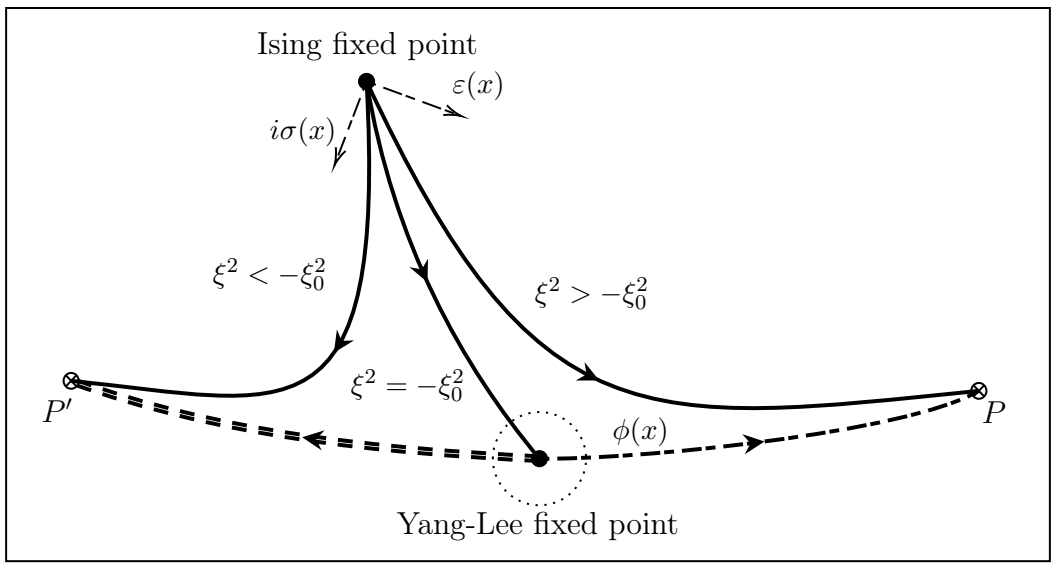
\begin{figure}[htb]
\centering
\begin{tikzpicture}[x=0.6pt,y=0.6pt,yscale=-1,xscale=1]
%uncomment if require: \path (0,925); %set diagram left start at 0, and has height of 925

%Shape: Rectangle [id:dp9099136009261399]
\draw   (1.56,556.78) -- (659.56,556.78) -- (659.56,905.78) -- (1.56,905.78) -- cycle ;
%Shape: Circle [id:dp21120689918883695]
\draw  [fill={rgb, 255:red, 0; green, 0; blue, 0 }  ,fill opacity=1 ] (224,603.11) .. controls (224,600.47) and (226.14,598.33) .. (228.78,598.33) .. controls (231.42,598.33) and (233.56,600.47) .. (233.56,603.11) .. controls (233.56,605.75) and (231.42,607.89) .. (228.78,607.89) .. controls (226.14,607.89) and (224,605.75) .. (224,603.11) -- cycle ;
%Flowchart: Summing Junction [id:dp7863763758608597]
\draw   (37.78,792) .. controls (37.78,789.24) and (39.92,787) .. (42.56,787) .. controls (45.19,787) and (47.33,789.24) .. (47.33,792) .. controls (47.33,794.76) and (45.19,797) .. (42.56,797) .. controls (39.92,797) and (37.78,794.76) .. (37.78,792) -- cycle ; \draw   (39.18,788.46) -- (45.93,795.54) ; \draw   (45.93,788.46) -- (39.18,795.54) ;
%Shape: Circle [id:dp41284840469388895]
\draw  [fill={rgb, 255:red, 0; green, 0; blue, 0 }  ,fill opacity=1 ] (333,841.11) .. controls (333,838.47) and (335.14,836.33) .. (337.78,836.33) .. controls (340.42,836.33) and (342.56,838.47) .. (342.56,841.11) .. controls (342.56,843.75) and (340.42,845.89) .. (337.78,845.89) .. controls (335.14,845.89) and (333,843.75) .. (333,841.11) -- cycle ;
%Flowchart: Summing Junction [id:dp533500561220515]
\draw   (610,798.33) .. controls (610,795.57) and (612.14,793.33) .. (614.78,793.33) .. controls (617.42,793.33) and (619.56,795.57) .. (619.56,798.33) .. controls (619.56,801.09) and (617.42,803.33) .. (614.78,803.33) .. controls (612.14,803.33) and (610,801.09) .. (610,798.33) -- cycle ; \draw   (611.4,794.8) -- (618.16,801.87) ; \draw   (618.16,794.8) -- (611.4,801.87) ;
%Curve Lines [id:da9899080412474484]
\draw [line width=1.5]    (228.78,603.11) .. controls (245.56,705) and (288.56,782) .. (337.78,841.11) ;
\draw [shift={(267.18,729.81)}, rotate = 245.66] [fill={rgb, 255:red, 0; green, 0; blue, 0 }  ][line width=0.08]  [draw opacity=0] (13.4,-6.43) -- (0,0) -- (13.4,6.44) -- (8.9,0) -- cycle    ;
%Curve Lines [id:da9706496572236483]
\draw [line width=1.5]  [dash pattern={on 3.75pt off 3pt on 7.5pt off 1.5pt}]  (337.78,841.11) .. controls (396.56,842) and (569.56,825) .. (614.78,798.33) ;
\draw [shift={(479.67,830.28)}, rotate = 532.65] [fill={rgb, 255:red, 0; green, 0; blue, 0 }  ][line width=0.08]  [draw opacity=0] (13.4,-6.43) -- (0,0) -- (13.4,6.44) -- (8.9,0) -- cycle    ;
%Curve Lines [id:da9871581013468786]
\draw [line width=1.5]  [dash pattern={on 5.63pt off 4.5pt}]  (337.7,842.61) .. controls (234.33,837.49) and (136.18,827.46) .. (42.05,793.41)(337.85,839.61) .. controls (234.78,834.51) and (136.93,824.54) .. (43.07,790.59) ;
\draw [shift={(188.18,827.61)}, rotate = 368.28999999999996] [fill={rgb, 255:red, 0; green, 0; blue, 0 }  ][line width=0.08]  [draw opacity=0] (13.4,-6.43) -- (0,0) -- (13.4,6.44) -- (8.9,0) -- cycle    ;
%Curve Lines [id:da9110353888104439]
\draw [line width=1.5]    (228.78,603.11) .. controls (335.56,832) and (402.56,820) .. (614.78,798.33) ;
\draw [shift={(375.5,791.78)}, rotate = 205.57] [fill={rgb, 255:red, 0; green, 0; blue, 0 }  ][line width=0.08]  [draw opacity=0] (13.4,-6.43) -- (0,0) -- (13.4,6.44) -- (8.9,0) -- cycle    ;
%Curve Lines [id:da08099931305160002]
\draw [line width=1.5]    (228.78,603.11) .. controls (241.56,836) and (179.56,806) .. (42.56,792) ;
\draw [shift={(208.2,774.24)}, rotate = 302.97] [fill={rgb, 255:red, 0; green, 0; blue, 0 }  ][line width=0.08]  [draw opacity=0] (13.4,-6.43) -- (0,0) -- (13.4,6.44) -- (8.9,0) -- cycle    ;
%Straight Lines [id:da10686977207524384]
\draw  [dash pattern={on 3.75pt off 3pt on 7.5pt off 1.5pt}]  (228.78,603.11) -- (209.27,654.13) ;
\draw [shift={(208.56,656)}, rotate = 290.92] [color={rgb, 255:red, 0; green, 0; blue, 0 }  ][line width=0.75]    (10.93,-3.29) .. controls (6.95,-1.4) and (3.31,-0.3) .. (0,0) .. controls (3.31,0.3) and (6.95,1.4) .. (10.93,3.29)   ;
%Straight Lines [id:da45854093578010735]
\draw  [dash pattern={on 3.75pt off 3pt on 7.5pt off 1.5pt}]  (228.78,603.11) -- (286.68,624.31) ;
\draw [shift={(288.56,625)}, rotate = 200.11] [color={rgb, 255:red, 0; green, 0; blue, 0 }  ][line width=0.75]    (10.93,-3.29) .. controls (6.95,-1.4) and (3.31,-0.3) .. (0,0) .. controls (3.31,0.3) and (6.95,1.4) .. (10.93,3.29)   ;
%Shape: Circle [id:dp872359657052209]
\draw  [dash pattern={on 0.84pt off 2.51pt}] (308.39,841.11) .. controls (308.39,824.88) and (321.55,811.72) .. (337.78,811.72) .. controls (354.01,811.72) and (367.17,824.88) .. (367.17,841.11) .. controls (367.17,857.34) and (354.01,870.5) .. (337.78,870.5) .. controls (321.55,870.5) and (308.39,857.34) .. (308.39,841.11) -- cycle ;

% Text Node
\draw (166,624) node [anchor=north west][inner sep=0.75pt]    {$i\sigma ( x)$};
% Text Node
\draw (281,595) node [anchor=north west][inner sep=0.75pt]    {$\varepsilon ( x)$};
% Text Node
\draw (158,570) node [anchor=north west][inner sep=0.75pt]  [font=\large] [align=left] {Ising fixed point};
% Text Node
\draw (246,873) node [anchor=north west][inner sep=0.75pt]  [font=\large] [align=left] {Yang-Lee fixed point};
% Text Node
\draw (217,783) node [anchor=north west][inner sep=0.75pt]    {$\xi ^{2} =-\xi _{0}^{2}$};
% Text Node
\draw (130,713) node [anchor=north west][inner sep=0.75pt]    {$\xi ^{2} < -\xi _{0}^{2}$};
% Text Node
\draw (333,725) node [anchor=north west][inner sep=0.75pt]    {$\xi ^{2}  >-\xi _{0}^{2}$};
% Text Node
\draw (614.78,803.33) node [anchor=north west][inner sep=0.75pt]  [font=\large]  {$P$};
% Text Node
\draw (22,801.33) node [anchor=north west][inner sep=0.75pt]  [font=\large]  {$P'$};
% Text Node
\draw (382,813) node [anchor=north west][inner sep=0.75pt]    {$\phi ( x)$};
\end{tikzpicture}
\caption{The topology of the RG flow at pure imaginary $h$. Critical and non-critical RG fixed points are given by bullets and crossed circles respectively. Some RG trajectories originated from Ising fixed point are shown with solid arrowed lines. The trajectories are labeled by $\xi^2$, and $\xi^2 = -\xi^2_0$ denotes the massless flow down to YL fixed point. Single and double dashed lines are showing the RG flows from the Yang-Lee fixed point to non-critical fixed points $P$ and $P'$, with positive and negative $\lambda$ respectively.}
\label{RGflowsPicture}
\end{figure}

The transition at pure imaginary $h$ described in the previous subsection, as well as the diagram in Fig.\ref{complexxiphasediagram}, has clear interpretation in terms of the Renormalization Group (RG) flow. Since RG flow represents just the change of the overall scale, the scaling parameter $\xi$ labels the RG trajectories. Because the change of sign of $h$ in \eqref{ift} can be compensated by the field transformation $\sigma(x) \to -\sigma(x)$, such transformation would act invariantly on the theories with pure imaginary $h = ig$. The topology of the RG flow with pure imaginary $h$ (real $g$) is shown schematically in Fig.\ref{RGflowsPicture}. The point $\xi^2=-\xi_0^2$ in Fig.\ref{complexxiphasediagram} represents the RG flow from the Ising fixed point $\mathcal{A}_\text{ICFT}$ down to the Yang-Lee fixed point $\mathcal{A}_\text{YLCFT}$. The two trajectories which originate at the Yang-Lee fixed point and flow to the
non-critical fixed points\footnote{As usual, "non-critical" refers to a fixed point with zero correlation length $R_c=0$ \cite{wilson1974renormalization}. The points $P$ and $P'$ in Fig.\ref{RGflowsPicture} are understood as follows. At $P$ the field $\phi$ is essentially frozen at certain real value ${\bar\phi}$, whereas $P'$ represents a superposition of the states with $\phi$ concentrated near two complex conjugate values.} $P$ and $P'$ represent the Yang-Lee QFT \eqref{ylqft} with positive and negative $\lambda$, respectively. Close vicinity of the point $\xi^2=-\xi_0^2$ in Fig.\ref{complexxiphasediagram} consists of trajectories which originate at the Ising fixed point, quickly approach the neighborhood of the Yang-Lee fixed point but narrowly miss it, and after long stay close to $\mathcal{A}_\text{YLCFT}$ finally depart towards the non-critical fixed points, following closely the YL QFT trajectories. The last two stages of the RG evolution are responsible for the formation of the critical singularities at the Yang-Lee point in Fig.\ref{complexxiphasediagram}. The singularities of the thermodynamic and correlation characteristics of the IFT near the Yang-Lee criticality $\xi^2 = -\xi_0^2$ is then governed by the effective action
\begin{eqnarray}\label{aeff0}
\mathcal{A}_\text{eff} = \mathcal{A}_\text{YLCFT} + \lambda\int\phi(x) d^2 x +
\sum_i  a_i \int O_i(x) d^2 x
\end{eqnarray}
where the first two terms constitute the Yang-Lee QFT, Eq.\eqref{ylqft}, and the sum represent contributions from the infinite tower of irrelevant scalar operators from the fixed point CFT $\mathcal{A}_\text{YLCFT}$, introduced to capture the structure of the RG flow in the vicinity of the YL fixed point. The fields $O_i (x)$ are scalars of the scale dimensions $2\Delta_i$ with $\Delta_i >1$; correspondingly, the coupling constants $a_i$ in \eqref{aeff0} have mass dimensions $a_i \sim[\text{mass}]^{2-2\Delta_i}$. Below we will say more about the content of the irrelevant operators appearing in the expansion \eqref{aeff0}.

The effective action \eqref{aeff0} can be understood as the result of the RG flow from the vicinity of the Ising fixed point
to the neighborhood of the Yang Lee fixed point. More precisely, the theory \eqref{ift} with sufficiently small $\xi^2+\xi_0^2$
flows to
\begin{eqnarray}\label{CosmTerm}
\mathcal{A}_\text{IFT} \ \rightarrow \ f\,V + \mathcal{A}_\text{eff}\,,
\end{eqnarray}
which differs from \eqref{aeff0} by the "induced cosmological term"; here $V$ is the volume of the 2D space-time, and $f$ is $\xi^2$-dependent constant. This term does not affect the physical content of the theory apart from bringing in the regular term
$f(\xi^2)$ in Eq.\eqref{Fsing} below. We mention it here because in statistical mechanics it is the full specific free energy $F$, not just the singular terms in \eqref{Fsing} which is directly measurable (In fact, it takes quite an elaborate analysis to isolate the singular terms from the data, see \cite{fonseca2003ising}).

The effective theory describes the vicinity of the critical point
$\xi^2=-\xi_0^2$ of the Ising QFT \eqref{ift}, and the parameters $f$, ${\lambda}$ as well as all ${a}_i$ depend on the scaling parameter $\xi^2$. The functions $f(\xi^2)$, ${\lambda}(\xi^2)$, and ${a}_i(\xi^2)$ are analytic at the point $-\xi_0^2$ and in some domain of $\xi^2$ surrounding this point \cite{wilson1974renormalization}; these parameters enjoy the convergent power series expansions
\begin{eqnarray}\label{lambdaexp}
&&{\lambda}(\xi^2) = (\xi^2+\xi_0^2)\,\lambda_1 + (\xi^2+\xi_0^2)^2\,\lambda_2 + ...\label{lambdaexp}\\
&&f(\xi^2) = f_0 + (\xi^2+\xi_0^2) \,f_1 + (\xi^2+\xi_0^2)^2 \,f_2 + ...\label{fexp}
\end{eqnarray}
and
\begin{eqnarray}\label{aexp}
{a}_{i}(\xi^2) = a_{i,0} + (\xi^2+\xi_0^2)\,a_{i,1} + (\xi^2+\xi_0^2)^2\,a_{i,2} + ...
\end{eqnarray}
The condition that $\lambda(-\xi_0^2)=0$ in the effective theory \eqref{aeff0} associated with the QFT \eqref{ift} is tautologically equivalent to the statement that $\xi^2 = -\xi_0^2$ is the critical point. One of the objectives of the present work is to give estimates of the most important of the coefficients $\lambda_k$ and $a_{i,k}$.

The effective action \eqref{aeff0} generates the singular expansions of physical quantities in fractional powers of $\xi^2+\xi_0^2$. Thus, the singular part $F_\text{sing}$ of the specific free energy, defined as
\begin{eqnarray}\label{Fsing}
F = f + F_\text{sing}\,,
\end{eqnarray}
and the mass $M(\xi^2)$, admit the expansions
\begin{eqnarray}\label{Fexpansion0}
&&F_\text{sing}(\xi^2) = B_\text{YL}\,[{\lambda(\xi^2)}]^\frac{5}{6} + a_1(\xi^2)\, B_1\, [\lambda(\xi^2)]^\frac{5\Delta_1}{6} +  \text{higher terms} \,,\\
&&M(\xi^2)\ \  =\  C_\text{YL}\,[{\lambda}(\xi^2)]^{\frac{5}{12}}+{a}_1(\xi^2)\,C_1\,\left[\lambda(\xi^2)\right]^{\frac{5}{12}(2\Delta_1-1)} +\text{higher terms}\,,\label{Mexpansion0}
\end{eqnarray}
where we assumed that $O_1(x)$ is the lowest of the irrelevant operators in \eqref{aeff0}. The constants
$B_\text{YL}, B_1, C_\text{YL}, C_1$, as well as similar coefficients in the higher terms are computable, in principle, from the YLQFT \eqref{ylqft}. In particular, $B_\text{YL} = -\frac{\sqrt{3}}{12} C_{\text{YL}}^2$, where $C_\text{YL}$ is given in  Eq.\eqref{cyl}. Below we will say more about $B_1, C_1$ and some higher coefficients in the singular expansion \eqref{Mexpansion0}.

\subsection*{Finite Size Spectrum and TFFSA}

\begin{figure}
\centering
\begin{tikzpicture}[x=0.5pt,y=0.5pt,yscale=-1,xscale=1]
%uncomment if require: \path (0,1584); %set diagram left start at 0, and has height of 1584

%Shape: Ellipse [id:dp7677720415681593]
\draw   (256.56,1308.44) .. controls (256.56,1297.4) and (289.56,1288.44) .. (330.28,1288.44) .. controls (370.99,1288.44) and (404,1297.4) .. (404,1308.44) .. controls (404,1319.49) and (370.99,1328.44) .. (330.28,1328.44) .. controls (289.56,1328.44) and (256.56,1319.49) .. (256.56,1308.44) -- cycle ;
%Straight Lines [id:da054186308326037214]
\draw    (256.56,1308.44) -- (257.56,1433.67) ;
\draw [shift={(257.06,1371.06)}, rotate = 89.54] [fill={rgb, 255:red, 0; green, 0; blue, 0 }  ][line width=0.08]  [draw opacity=0] (10.72,-5.15) -- (0,0) -- (10.72,5.15) -- (7.12,0) -- cycle    ;
%Straight Lines [id:da22672497637054279]
\draw    (404,1308.44) -- (404.56,1432.89) ;
%Curve Lines [id:da09069980724145754]
\draw    (257.56,1433.67) .. controls (268.11,1458.89) and (400.11,1458.89) .. (404.56,1432.89) ;
\draw [shift={(332.3,1452.49)}, rotate = 180.68] [fill={rgb, 255:red, 0; green, 0; blue, 0 }  ][line width=0.08]  [draw opacity=0] (10.72,-5.15) -- (0,0) -- (10.72,5.15) -- (7.12,0) -- cycle    ;
%Curve Lines [id:da9724095156735166]
\draw  [dash pattern={on 0.84pt off 2.51pt}]  (257.56,1433.67) .. controls (267.11,1406.89) and (391.11,1406.89) .. (404.56,1432.89) ;

% Text Node
\draw (324,1457.44) node [anchor=north west][inner sep=0.75pt]  [font=\large]  {$\tx$};
% Text Node
\draw (222,1358.44) node [anchor=north west][inner sep=0.75pt]  [font=\large]  {$\ty$};
\end{tikzpicture}
\caption{Space-time cylinder, with $\tx \sim \tx + R$}
\label{cylinder}
\end{figure}
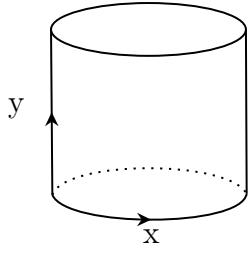

Technically, most of our analysis will be in terms of the energy spectrum of the theory in the finite size geometry. We consider the theory \eqref{ift} in the geometry of a long Euclidean cylinder, with the "spatial" coordinate $\tx$ compactified on a circle of the circumference $R$, $\tx\sim \tx +R$, while the complimentary Cartesian coordinate $\ty$ playing the role of imaginary time, see Fig.\ref{cylinder}. At finite $R$ the energy spectrum is discrete, and generally non-degenerate. We denote $|n \rangle_R, \ n=0,1,2,3,...$ the consecutive eigenstates of the finite-size Hamiltonian of \eqref{ift} with the spatial momentum $P_n=0$, and assume the standard normalization
\begin{eqnarray}\label{statenorm}
_R\langle \; n | n' \rangle_R = \delta_{n,n'}\,.
\end{eqnarray}
The $R$ dependence of the corresponding energy eigenvalues $E_n(R)$ will be the main instrument of our analysis.
%\begin{figure}[ht]
%\centering
%\includegraphics[width=4cm]{cylinder.eps}
%\caption{Space-time cylinder } \label{**}
%\end{figure}

The RG flow can be traced by going from short distances, $R\to 0$, to long distances $R \gg M^{-1}$. While
\begin{eqnarray}\label{enuv}
E_n(R) \ \to\ - \frac{C_{n}^{(\text{UV})}}{12}\,\frac{2\pi}{R} \qquad \text{as}\quad R\to 0
\end{eqnarray}
with the constants $C_{n}^{(\text{UV})}$ determined by the UV fixed point CFT, at large $R$ the levels $E_n(R)$ behave as
\begin{eqnarray}\label{enir}
E_n(R)=F R + {\cal E}_n(R)
\end{eqnarray}
where the coefficient $F$ is identified with the bulk vacuum energy density, and ${\cal E}_n(R)$ are bounded at large $R$.
The large $R$ behavior of ${\cal E}_n(R)$ depends on whether the RG flows to a non-critical or a critical IR fixed point.
At the critical point the the behavior is governed by the CFT associated with the IR fixed point,
\begin{eqnarray}
{\cal E}_n (R) \ \to\ - \frac{C_{n}^{(\text{IR})}}{12}\,\frac{2\pi}{R} \qquad \text{as}\quad R\to \infty
\end{eqnarray}
where the coefficients now depend on the central charge and dimensions of the IR CFT. Away from the critical point, when the correlation length $R_c = M^{-1}$ is finite, ${\cal E}_n(R)$ approach the finite-size spectrum of massive particles,
the details being determined by mass spectrum and the S-matrix of the massive QFT. Thus, the lowest two levels are
the vacuum state and the state of one particle at rest,
\begin{eqnarray}
&&{\cal E}_0(R) \ =\ O(e^{-M R})\label{e0ir}\\
&&{\cal E}_1(R) = M + O(e^{-\frac{\sqrt{3}}{2}M R})\label{e1ir}
\end{eqnarray}

At generic values of $\xi^2$ (including positive, negative, and even complex values of this parameter) one can compute the spectrum $E_n(R)$ using the so-called Truncated Free Fermion Space Approach (TFFSA) introduced in \cite{yurov1991truncated}, and further developed in \cite{fonseca2003ising}. It is a modification of the Truncated Conformal Space Approach of \cite{yurov1990truncated}, specifically designed to handle the IFT \eqref{ift}. It utilizes the fact that in the absence of the last term in \eqref{ift}, i.e. at $h=0$, the IFT reduces to the theory of free Majorana fermions
\begin{eqnarray}\label{ffqft}
\mathcal{A}_\text{FF}=\frac{1}{2\pi}\,\int\,\left[\psi{\bar\partial}\psi + {\bar\psi}\partial{\bar\psi}
+i m\,{\bar\psi}{\psi}\right]\,d^2 x
\end{eqnarray}
where $(\psi, {\bar\psi})$ are two components of the neutral fermi field, and $(\partial,{\bar\partial})=(\frac{\partial_\tx-i\partial_\ty}{2},\frac{\partial_\tx+i\partial_\ty}{2})$. It is of course the theory of free neutral fermi particles of the mass $|m|$, and the space of its states is the fermionic Fock space, spanned by the multi-particle states
\begin{eqnarray}\label{ffstates}
| p_1, p_2, ... , p_N\rangle
\end{eqnarray}
which are the eigenvectors of the Hamiltonian $H_\text{FF}$ of the theory \eqref{ffqft} with the eigenvalues
\begin{eqnarray}
E_\text{vac} + \sum_{i=1}^N \,\omega(p_i)\,, \qquad \omega(p)=\sqrt{m^2+p^2}\,.
\end{eqnarray}
If the coordinate $\ty$ along the cylinder in Fig.\ref{cylinder} is chosen to be the (Euclidean) time, the momenta $p_i$
are quantized, $p_i = \frac{2\pi\,k_i}{R}$, where $k_i$ are integers or half-integers, depending on whether we take periodic or anti-periodic boundary conditions for the fermi field $\psi$, $\psi(\tx,\ty)=\psi(\tx+R,\ty)$ (R-sector) or $\psi(\tx,\ty)=-\psi(\tx+R,\ty)$ (NS sector). There are certain conditions on the fermion number in each of these sectors, which depend on the sign of $m$. All details of the structure of the space of states of the spatially finite system \eqref{ffqft} can be found in Ref.\cite{fonseca2003ising} (see also \cite{mccoy2013two}), including expressions for the finite size ground state energies in each sector, $E_{0\, (R)}(R)$ and $E_{0\, (NS)}(R)$.

When a non-zero magnetic field $h$ is added, the Hamiltonian acquires an additional term
\begin{eqnarray}\label{hift}
H_\text{IFT} = H_\text{FF} + h\,\int_0^R\,\sigma(x)\,d\tx
\end{eqnarray}
where the integral in the second term is over the equal-time slice $x=(\tx, \ty=0)$, and $\sigma(x)$ is the
"spin field". In terms of the representation \eqref{ffqft}, the operator $\sigma(x)$ creates dislocation in the field $(\psi,{\bar \psi})$ at the point $x$ ($(\psi,{\bar \psi})$ changes to $(-\psi,-{\bar \psi})$ when one goes around the point $x$), and hence it intertwines the R- and NS- sectors. All the matrix elements of $\sigma(x)$ between the Fock states \eqref{ffstates} are known in a closed form \cite{fonseca2003ising}, and the energy levels $E_n(R)$ of the full IFT \eqref{ift} can be found by diagonalizing the operator \eqref{hift}. To render this problem amendable to numerical analysis one needs to make finite-dimensional approximation for the space of states in which the operator \eqref{hift} acts. As in \cite{fonseca2003ising}, we truncate the space of states according to the condition
\begin{eqnarray}\label{TruncationLevelDefinition}
\sum_{i=1}^N |k_i| \leq 2L
\end{eqnarray}
where the number $L$ (integer for all admissible states \cite{fonseca2003ising}) is called the "truncation level"\footnote{The level roughly
correlates with the maximal energies of the admitted states, whereas this truncation is relatively easy to
implement. This truncation method is very similar to that used in TCSA \cite{yurov1990truncated}.}.
When $L$ is increased, the low-laying eigenvalues obtained by the numerical diagonalization of the truncated Hamiltonian \eqref{hift} stabilize and approximate well the exact eigenvalues of the full theory, as long as the spatial compactification size $R$ is not too large. The accuracy of the results can be judged by the
$L$-dependence of the eigenvalues in the truncated space. The results with $L=11,12,13$ are usually very stable for few lowest eigenvalues, for $|m|R\leq 10$. For larger $R$ the inaccuracy signified by the dependence on $L$ - the "truncation effects" - is still noticeable at $L=13$, while with larger truncation levels the numerical diagonalization becomes prohibitively difficult. The described procedure can be applied to \eqref{hift} with any complex $h$, but here we limit attention to the cases of real and pure imaginary $h$
(real $h^2$).

\subsection*{Results}

In this work we determine the most important parameters of the effective action \eqref{aeff0}. We give numerical estimates
of the leading coefficients in the expansions \eqref{lambdaexp},\eqref{fexp} (Eqs\eqref{lambdaestimate}, \eqref{f0estimate} below), as well as in the expansions \eqref{aexp} for the most important irrelevant couplings, see Eqs\eqref{alphaestimate},\eqref{betaestimate}. For the mass function $M(\xi^2)$ we give some detalization of the singular expansion \eqref{Mexpansion0}. We also confirm the expected analyticity of $M(\xi^2)$ on the whole complex $\xi^2$ plane with the branch cut along the real axis from $-\infty$ to $-\xi^2$, as shown in Fig.\ref{complexxiphasediagram}.

\section{Yang-Lee Criticality and Singular Expansion}

Here we describe the general structure of the effective action \eqref{aeff0} associated with the Yang-Lee
criticality, along with some exact results, notably the role played by the "TTbar" contributions.

\subsection*{Space of (scalar) fields in YL CFT}

As is well known, the space of local fields $\mathcal{F}_\text{YLCFT}$ of the minimal CFT $\mathcal{M}_{2/5}$ involves two irreducible representations of the Virasoro algebra with the central charge $c_{\text{YL}}=-\frac{22}{5}$,
\begin{eqnarray}\label{fylcft}
\mathcal{F}_\text{YLCFT} = \left(\mathcal{V}_0 \otimes {\bar{\cal V}}_0\right)\oplus
\left(\mathcal{V}_{-1/5} \otimes {\bar{\cal V}}_{-1/5}\right)
\end{eqnarray}
where ${\cal V}_\Delta$ denotes the irreducible Virasoro module with the lowest weight $\Delta$.
The first and the second term in the direct sum in \eqref{fylcft} consist of the Virasoro descendants of the two primary fields in this CFT, the identity field $I$ and the field $\phi$, respectively. The latter is a scalar, and its left and right Virasoro dimensions are $(-\frac{1}{5},-\frac{1}{5})$. Since only scalar fields can appear in \eqref{aeff0} (as the flow from Ising CFT down to YL CFT preserves the rotational symmetry), only the scalar fields, i.e. the descendants of the dimensions $(l,l)$ and $(-\frac{1}{5}+l, -\frac{1}{5}+l)$, enter the effective action \eqref{aeff0}; the positive integer $l$ represents the level of the descendant. The scale dimensions of the level $l$ descendants of $I$ or $\phi$ are equal to $2l$ or $-\frac{2}{5}+2l$, respectively. The fields which are the space-time derivatives of another local fields (i.e. the descendants generated by the Virasoro generators $L_{-1}$ and ${\bar L}_{-1}$) can be ignored, as they don't contribute to the bulk theory \eqref{aeff0}. The space of fields that can enter the effective action \eqref{aeff0} is therefore isomorphic to ${\hat{\cal V}}_0 \oplus {\hat{\cal V}}_{-1/5}$, where ${\hat{\cal V}}_\Delta$ stands for the factor spaces ${\cal V}_\Delta/L_{-1}{\cal V}_\Delta$. The numbers of the independent descendants ${\cal N}_I^{(l)}$ and ${\cal N}_\phi^{(l)}$ which may appear in \eqref{aeff0} are then computed as the coefficients of the $q$-expansions of $(1-q)\chi_0(q)+q$ and $(1-q) \chi_{-1/5}(q)$, respectively, where $\chi_\Delta(q)$ are characters of the irreducible Virasoro moduli at $c_{\text{YL}}=-\frac{22}{5}$. Table \ref{Tab:dimensions} shows these  multiplicities for few lowest levels.
\begin{table}[h]
\begin{center}
 \begin{tabular}{|c | c c c c c c c c c c c c c c c|}
 \hline
 $l$ & 0 & 1 & 2 & 3 & 4 & 5 & 6 & 7 & 8 & 9 & 10 & 11 & 12 & 13 & 14 \\
  \hline
 $\mathcal N^{(l)}_{I}$ & 1 & 0 & 1 & 0 & 0 & 0 & 1 & 0 & 1 & 0 & 1 & 0 & 2 & 0 & 2\\
 \hline
 ${\mathcal N}^{(l)}_{\phi}$ & 1 & 0 & 0 & 0 & 1 & 0 & 1 & 0 & 1 & 1 & 1 & 1 & 2 & 1 & 2\\
 \hline
\end{tabular}
\caption{Dimensionalities ${\cal N}_0^{(l)}$ and ${\cal N}_{-1/5}^{(l)}$ of the level $l$ subspaces in ${\hat{\cal V}}_0$ and ${\hat{\cal V}}_{-1/5}$}
\label{Tab:dimensions}
\end{center}
\end{table}
We see that at low levels the nonzero entries are relatively sparse. This is related to the fact that at $c_{\text{YL}}=-\frac{22}{5}$ the Virasoro moduli have additional null vectors, which must be factored out of the irreducible representations. Thus, the module $\mathcal{V}_{-1/5}$ has two independent null vectors
\begin{eqnarray}\label{nullphi}
\left[L_{-2} - \frac{5}{2} L_{-1}^2\right]\phi=0\,,\qquad \left[L_{-3}-\frac{10}{9} L_{-1} L_{-2} + \frac{25}{36} L_{-1}^3\right]\phi=0\,.
\end{eqnarray}
Likewise, the irreducible module $\mathcal{V}_0$ is obtained by factoring out the null-vectors
\begin{eqnarray}\label{nullI}
L_{-1}I=0\,, \qquad \left[L_{-4}-\frac{5}{3} L_{-2}^2\right]I=0\
\end{eqnarray}
together with all their descendants.

Let us briefly review few lowest irrelevant descendants. The lowest nontrivial scalar descendant of the identity operator is $L_2 {\bar L}_{-2}I$, alternatively known as $T{\bar T}$. Thus, the field $T{\bar T}$ generally brings in the least irrelevant contribution to the effective action \eqref{aeff0}. Adding this operator generates the so-called "TTbar deformation", which allows for much analytic control over its contribution to finite-size energy levels associated with \eqref{aeff0} (see below). In view of
the null-vector equations \eqref{nullI}, the descendants of $I$ at the levels $3,4,5$ are all total derivatives of the $T{\bar T}$, that is why the slots $l=3,4,5$ in the first raw of Table \ref{Tab:dimensions} are empty. The next nonzero entry in that row appears at the level 6; the corresponding descendant $L_{-3}^2 {\bar L}_{-3}^2 I$ has the scale dimension 12. In Sec.5 we will say more about the higher descendants of $I$ at $l=8,10,12,\dots$.

The first nontrivial descendant of $\phi$ appear at the level 4. In what follows we use the notation
\begin{eqnarray}\label{Xidef}
\Xi (x) = \left(L_{-4}-\frac{625}{624} L_{-1}^4\right)\left({\bar L}_{-4}-\frac{625}{624} {\bar L}_{-1}^4\right)\phi(x)\,.
\end{eqnarray}
for the scalar level 4 quasi-primary descendant\footnote{The terms with $L_{-1}$ and ${\bar L}_{-1}$ are total derivatives, and play no role in the effective action \eqref{aeff1} below. However, using the quasi-primary form \eqref{Xidef} simplifies calculation of
its matrix elements and correlation function.}. Its scale dimension $2\Delta_{\Xi} = \frac{38}{5} = 7.6$ is greater than the dimension of $T{\bar T}$ but lower than the dimensions of the higher descendants of the identity. The next non-derivative scalar descendant of $\phi$ is at the
level 6; its scale dimension is $\frac{58}{5} = 11.6$.

\subsection*{Effective action and perturbative analysis}

In this work we disregard all irrelevant operators with the mass dimension greater than $38/5$ in \eqref{aeff0}, considering the effective action
\begin{eqnarray}\label{aeff1}
&&\mathcal{A}_\text{eff} = \mathcal{A}_\text{YLCFT} + \lambda\int\phi(x) \, d^2 x + \frac{\alpha}{\pi^2} \,\int T{\bar T}(x) \, d^2 x +
%\qquad\qquad \qquad\qquad\qquad \\
%&&\qquad\qquad\qquad\qquad\qquad\qquad
 {\frac{\beta}{2\pi}}\, \int \Xi(x) \, d^2 x \,,
\end{eqnarray}
where $\alpha$ and $\beta$ are coupling constants with negative mass dimensions, $\alpha \sim \ [\text{mass}]^{-2}\,,\ \beta \sim\  [\text{mass}]^{-28/5}$ \footnote{Note our definition of the coupling constant $\alpha$ here agrees with the notations in \cite{smirnov2017space}, but differs by the factor $1/4$ from the eponymous parameter defined in \cite{fonseca2003ising}.}. In IFT they depend on the scaling parameter $\xi^2$, and admit convergent expansions in the powers of $\xi^2+\xi_0^2$,
\begin{eqnarray}
&&\alpha(\xi^2) = \alpha_0 + (\xi^2+\xi_0^2) \alpha_1 + ...\label{alphaxi}\,,\\
&&\beta(\xi^2) = \beta_0 + (\xi^2+\xi_0^2) \beta_1 + ...\label{betaxi}\,.
\end{eqnarray}

Since we regard \eqref{aeff1} as the perturbation of the full theory YLQFT \eqref{ylqft}, not just the CFT point, let us briefly comment on how the fields $T{\bar T}(x)$ and $\Xi(x)$ are defined in YLQFT away from the CFT point $\lambda=0$. The field $T{\bar T}(x)$ is universally defined in generic 2D QFT in terms of its energy-momentum tensor, see Ref.\cite{smirnov2017space}\footnote{Let us stress that here we define $T{\bar T}$ in terms of the energy-momentum tensor $T_{\mu\nu}$ of the effective theory \eqref{aeff0}, which differs from the energy-momentum tensor of the full RG flow \eqref{ift} by the "cosmological term", $T_{\mu\nu}^{\text{IFT}} = T_{\mu\nu} + f\,g_{\mu\nu}$, see Eq.\eqref{CosmTerm}. The difference between the flow generated by our $T{\bar T}$ and $T{\bar T}^{\text{IFT}}$ amounts to trivial scale renormalization.}. The field $\Xi(x)$ is the only scalar field of the dimension $\frac{38}{5}$ which is not a derivative of other local field, therefore the Eq.\eqref{Xidef} defines this field in the off-critical theory \eqref{ylqft} uniquely, up to the overall normalization and the derivative terms (see \cite{Zamolodchikov:1987zf}).  The latter ambiguity can be fixed by imposing the normalization conditions
\begin{eqnarray}\label{Xinorm}
\langle \Xi(x)\Xi(x')\rangle_\text{YLFT}\,|x-x'|^{\frac{76}{5}}\ \to\ N_\Xi^2\,, \qquad \langle \Xi(x)\phi(x')\rangle_\text{YLQFT}\,|x-x'|^\frac{36}{5}\ \to\ 0\,,
\end{eqnarray}
as $|x-x'| \to 0$. {Here $N_\Xi=\frac{3 \cdot 8803 }{2^{5/2} \cdot 5\cdot 13}$ is the norm of the CFT state associated with $\Xi$, and the second condition in \eqref{Xinorm} reflects our choice of the derivative terms in \eqref{Xidef}, which makes it a qusi-primary field in the CFT limit.}

Of course, the perturbation theory in the couplings $\alpha$ and $\beta$ is non-renormalizable. The perturbative calculations beyond the leading orders require introducing infinitely many largely undetermined counterterms, making the results ambiguous. However, the operator $T{\bar T}$ is special. This operator generates the so-called TTbar deformation \cite{smirnov2017space}\cite{Cavaglia:2016oda}, where the expansion in $\alpha$ is not only well defined, but for some quantities can be computed in a closed form (see the subsection below for some details). The TTbar deformation faithfully reproduces the perturbation series in $\alpha$ for \eqref{aeff1} up to the order $\alpha^5$ (the term $\sim \alpha^6$ competes dimension-wise with the contribution of the operator $L_{-3}^2{\bar L}_{-3}^2 I$, which is disregarded in \eqref{aeff1}, but may be present in \eqref{aeff0}).

Higher orders in $\beta$ are difficult even to define, and here we limit attention to the leading contributions $\sim \beta$ from the operator $\Xi$ in \eqref{aeff1}, which of course is well defined. Moreover, the TTbar deformation reproduces unambiguously
the terms $\sim \alpha\beta$ (again, the terms $\alpha^2 \beta$ and higher in $\alpha$ compete with the contributions from the
higher descendants of $\phi$ not accounted in \eqref{aeff1}).

As was mentioned in the Introduction, the Yang-Lee QFT \eqref{ylqft} is integrable. Moreover, the TTbar deformation of the integrable theory is integrable as well \cite{smirnov2017space}. On the other hand, the IFT \eqref{ift} at generic values of $\xi^2$, including neighborhood of the Yang-Lee critical point, is not integrable\footnote{The massless flow at $\xi^2=-\xi_0^2$, although converging to the integrable CFT in both the ultraviolet and infrared limits, is not integrable at all scales.}. Therefore, one expects that some irrelevant operators in \eqref{aeff0} break integrability.  The significance of the operator $\Xi$ in \eqref{aeff1} is that it is the lowest dimensional term which does that.

\subsubsection*{TTbar deformation}

As was mentioned in the previous paragraph, adding the lowest irrelevant term $\sim \int T{\bar T}(x) d^2 x$ in \eqref{aeff1} can be understood in terms of the "TTbar deformation" of the Yang-Lee QFT
\eqref{ylqft}. Generally, the TTbar deformation of a given theory ${\cal A}^{(0)}$ is defined by the
flow equation \cite{smirnov2017space}\cite{Cavaglia:2016oda}
\begin{eqnarray}\label{ttbarflow}
\frac{d}{d\alpha}{\cal A}^{(\alpha)} = \frac{1}{\pi^2}\,\int (T{\bar T})^{(\alpha)} (x) d^2 x
\end{eqnarray}
where $\alpha$ is the deformation parameter, and
$(T{\bar T})^{(\alpha)}(x)$ is a scalar local operator of exact dimension 4 built from the components of the
energy-momentum tensor of the deformed theory ${\cal A}^{(\alpha)}$, as is explained in Ref.\cite{zamolodchikov2004expectation}).
One can develop the solution of the flow equation \eqref{ttbarflow} as the power series in the deformation parameter $\alpha$. In the leading order this generates the term $\sim \int T{\bar T}(x)\, d^2 x$, as in \eqref{aeff1}. The higher orders in $\alpha$ bring in a string of the operators of higher dimensions, all
descendants of the identity operator ${I}$, i.e. belonging to the first row in the Table \ref{Tab:dimensions}\footnote{This is literally true in the case when the undeformed theory ${\cal A}^{(0)}$ is a CFT. In general case the operators are more complicated, but still can be understood as "deformations" of the corresponding descendants of ${I}$.}.

The TTbar deformation of a given QFT is "solvable", in the sense that some important quantities of the deformed theory can be found in a closed form in terms of the corresponding quantities in the undeformed theory. This in particular concerns the finite-size energy levels $E(R)$, which are uniquely determined by the finite-size energies of the undeformed theory ${\cal A}^{(0)}$. Below we consider only the states of the finite-size system (Fig.\ref{cylinder}) having zero spatial momentum $P_\tx=0$; in this case the relation is particularly simple. Given an energy level $E^{(0)}(R)$ of the undeformed theory ${\cal A}^{(0)}$, the associated level $E^{(\alpha)}(R)$ of ${\cal A}^{(\alpha)}$ is expressed as
\begin{eqnarray}\label{ealpha}
E^{(\alpha)}(R)=E^{(0)}\left(R-\alpha\, E^{(\alpha)}(R)\right)\,.
\end{eqnarray}
A simple consequence of \eqref{ealpha} is the $\alpha$-dependence of the vacuum energy density $F^{(\alpha)}$ in
Eq.\eqref{enir}, and the mass $M^{(\alpha)}$
\begin{eqnarray}\label{falpha}
F^{(\alpha)}=\frac{F^{(0)}}{1+\alpha F^{(0)}}\,, \qquad M^{(\alpha)}=\frac{M^{(0)}}{1+\alpha F^{(0)}} = M^{(0)}\,(1-\alpha F^{(\alpha)})\,.
\end{eqnarray}

\subsection*{Singular expansions near YL critical point}

The irrelevant operators in \eqref{aeff1} are responsible for the subleading terms in the singular expansions of the thermodynamic and correlation functions in fractional powers of $\xi^2+\xi_0^2$. Generally, for an effective action \eqref{aeff0}, the mass $M$ admits expansion in the irrelevant couplings $a_i$
\begin{eqnarray}\label{mexp1}
M=M_\text{YL} + \sum_i C^{(i)}\, a_i \,\left[M_\text{YL}\right]^{2\Delta_i-1} + \sum_{ij} \,C^{(ij)} \,
a_i a_j \,\left[M_\text{YL}\right]^{2\Delta_i +2\Delta_j -3} + ...
\end{eqnarray}
where $M_\text{YL}$ is the mass \eqref{myl} of the theory \eqref{ylqft}, and $C^{(i)}$, $C^{(ij)}$, ..., are numerical coefficients.
The coefficients $C^{(i)}$ at the leading order are related in a simple way to the diagonal matrix elements of the operators $O_i$ between the one-particle states\footnote{By Lorentz invariance, the matrix elements do not depend on $\theta$, but depend on the normalization of the states. We assume the standard normalization of the particle states, $\langle \theta_1 |
\theta_2\rangle = 2\pi\,\delta(\theta_1-\theta_2)$.},
\begin{eqnarray}
C^{(i)} \,M^{2\Delta_i} = \langle \theta | O_i (0) | \theta\rangle\,,
\end{eqnarray}
in the YLQFT \eqref{ylqft}. In general, the higher orders are largely undetermined. The perturbation theory in the couplings $a_i$ in \eqref{aeff0} is non-renormalizable, with all the usual problems related to the presence of an infinite number of ambiguous counterterms, and the coefficients $C$ in \eqref{mexp1} beyond the linear order are not uniquely determined through the perturbation theory. However, the expansion in $\alpha$ in \eqref{aeff1} constitutes a notable exception, as was explained in the previous Subsection. The coefficients of the $\alpha$-expansion are not only well defined, but can be computed in closed form. With the effective action \eqref{aeff1} the mass $M$ expands as follows
\begin{eqnarray}\label{Mexpansion1}
M = M_\text{YL}\,(1-\alpha F_\text{sing})\left(1+\frac{\beta}{2\pi}\, m_\Xi \,M_\text{YL}^{28/5} + O(M_\text{YL}^{48/5})\right)
\end{eqnarray}
where $M_\text{YL}$ is the mass of the YLQFT, Eq.\eqref{myl}, and $F_\text{sing}$ is the vacuum energy density of the
effective theory \eqref{aeff1}; it in turn expands as
\begin{eqnarray}\label{Fexpansion1}
F_\text{sing} = \frac{F_\text{sing}^{(0)}}{1+\alpha F_\text{sing}^{(0)}}\,,
\qquad F_\text{sing}^{(0)} = f_\text{YL}\,M_\text{YL}^2 + \frac{\beta}{2\pi}\,f_\Xi\,M_\text{YL}^{38/5} + O(M^{58/5})
\end{eqnarray}
($f_\text{YL}=-\sqrt{3}/12$). These expansions take into account the $T{\bar T}$ flow equations \eqref{falpha}, as well as the leading term in $\beta$. The numerical coefficients $m_\Xi$ and $f_\Xi$ are given by the diagonal matrix elements of the operator $\Xi$,
\begin{eqnarray}\label{Xielements}
m_\Xi\,M_\text{YL}^{38/5} = \langle\theta | \Xi(0)| \theta\rangle\,, \qquad f_\Xi\,M^{38/5}_\text{YL} = \langle 0 | \Xi(0) | 0\rangle\,,
\end{eqnarray}
in the YLQFT \eqref{ylqft}. The higher terms omitted in \eqref{Mexpansion1}. \eqref{Fexpansion1} can come from the contributions
of the higher irrelevant operators ($L_{-2}^3\bar L_{-2}^3 I$, $L_{-6}{\bar L}_{-6}\phi$, etc), neglected in \eqref{aeff1}, as well as the higher-order terms in $\beta$ and higher couplings. Since
\begin{eqnarray}\label{mylexpansion}
M_\text{YL} = M_\text{YL}(\xi^2) = C_\text{YL}\,\left[\lambda(\xi^2)\right]^{5/12} = C_\text{YL}\,(\xi^2+\xi_0^2)^{5/12}\, \left[\lambda_1 +
\lambda_2\,(\xi^2+\xi_0^2) + ...\right]^{5/12}
\end{eqnarray}
the expansion \eqref{Mexpansion1} translates into the singular expansion of $M(\xi^2)$ near the Yang-Lee critical point
\begin{eqnarray}\label{Mexpansion}
M(\xi^2) = (\xi^2+\xi_0^2)^{5/12}\,\left[b_0 + b_1\,(\xi^2+\xi_0^2) + c_0\,(\xi^2+\xi_0^2)^{5/6} + ...\right]
\end{eqnarray}
with the coefficients
\begin{eqnarray}\label{Mexpcoeffs}
b_0=C_\text{YL} \,\lambda_1^{5/12}\,, \quad b_1 = \frac{5}{12}\, C_{\text{YL}} \, \lambda_2 \, \lambda_1^{-7/12}\,,
\qquad c_0=-\alpha_0\,f_\text{YL}\,b_0^3 \,.
\end{eqnarray}

\section{Finite size spectrum at the Yang-Lee point}

In this section we shall discuss properties of the finite-size energy spectrum, i.e. the eigenvalues $E_n(R)$ of the Hamiltonian \eqref{hift}, on the cylinder geometry of Fig.\ref{cylinder}, at the Yang-Lee critical point $\xi^2=-\xi_0^2$.
For this value of $\xi^2$, the IFT \eqref{ift} describes the massless RG flow from the Ising fixed point down to the Yang-Lee fixed point (see Fig.\ref{RGflowsPicture}). Correspondingly, the $R\to 0$ limit of the levels $E_n(R)$ is determined by the Ising CFT ${\cal M}_{3/4}$, whereas their $R\to\infty$ behavior is controlled by the Yang-Lee CFT ${\cal M}_{2/5}$. Here we are specifically interested in the large $R$
expansions of the levels $E_n(R)$, whose leading terms are given by the eigenvalues of the operator
\begin{eqnarray}\label{hircft}
f_0\,R + \frac{2\pi}{R}\,{\bf H}_\text{YLCFT}\,,
\end{eqnarray}
where the asymptotic slope {$f_0=f(-\xi_0^2)$} is the vacuum energy density of \eqref{ift} at the Yang-Lee critical point,and ${\bf H}_\text{YLCFT}$ is the Hamiltonian of the YLCFT on the cylinder in Fig.\ref{cylinder} with $R=2\pi$.

Let us briefly recall the structure of the space of states ${\cal H}_\text{YLCFT}$ and the spectrum of
${\bf H}_\text{YLCFT}$, in order to fix the notations.

\subsection*{Spectrum of YLCFT}

By the standard operator-state correspondence of CFT, the space of states ${\cal H}_\text{YLCFT}$ of the YLCFT is isomorphic to the space \eqref{fylcft}, where ${\cal V}_\Delta$ (${\bar{\cal V}}_\Delta)$ stands for the irreducible lowest weight module over the left (right) Virasoro algebra with the lowest weight $\Delta$. The operators
\begin{eqnarray}\label{cylinderL}
 {\bf L}_n = - \frac{R}{2\pi}\,\int_{0}^R\,\frac{dz}{2\pi}\,T(z)\,e^{-\frac{2\pi i}{R} n z} + \frac{c}{24}\delta_{n,0},\
 {\bf{\bar L}}_n = - \frac{R}{2\pi}\,\int_{0}^R\,\frac{dz}{2\pi}\,{\bar T}(z)\,e^{-\frac{2\pi i}{R} n {\bar z}} + \frac{c}{24}\delta_{n,0}
\end{eqnarray}
where $z=\tx+i\ty$ and ${\bar z}=\tx-i\ty$ are the standard Cartesian coordinates on the cylinder in Fig.\ref{cylinder}, form the two commuting copies of the Virasoro algebra, with the commutators $\left[{\bf L}_n, {\bf L}_m\right] = (n-m)\,{\bf L}_{n+m} + \frac{c}{12}\,n(n^2-1)\,\delta_{n+m,\,0}$\,, and similarly for the ${\bar {\bf L}}$'s. Here $c$ is the central charge, which in this case takes the value $-22/5$ \cite{Cardy:1989fw}. We use bold face notations for these operators to distinguish them from the Virasoro generators $L_n$ acting on the space of fields \eqref{fylcft}, which are defined in terms of the integrals over small contours encircling the insertion point\footnote{The operators $L_n$ in \eqref{fylcft} are defined, as
$$
L_n O(z_0,\zb_0) = \oint_{{\cal C}_{z_0}}\,T(z)\,(z-z_0)^{n+1}\,O(z_0,\zb_0)\,\frac{dz}{2\pi i}\,,
$$
with similar definition for $\bar L_n$. The integration contour ${\cal C}_{z_0}$ is encircling the insertion point $z_0$ ($\bar z_0$) clockwise (anticlockwise), whereas the contour in \eqref{cylinderL} goes around the cylinder.}.

Let us denote $| I \rangle$ and $| \phi \rangle$ the primary states corresponding to the primary fields $I$ (the identity operator) and $\phi(x)$, respectively, so that ${\bf L}_0 | I \rangle= {\bf{\bar L}}_0 | I \rangle =0$ and ${\bf L}_0 | \phi \rangle = {\bar{\bf L}}_0 | \phi \rangle = -\frac{1}{5} | \phi \rangle$, with the standard CFT normalizations $\langle I | I \rangle = \langle \phi | \phi \rangle = 1$\footnote{We assume $| \phi \rangle=\lim_{y\to -\infty}\,i\phi(x,y)| I \rangle$ on the cylinder in Fig.\ref{cylinder}.}. The space ${\cal H}_\text{YLCFT}$ consists of these two primaries along with all their Virasoro descendants.

The finite-size Hamiltonian of YLCFT (with $R=2\pi$) is
\begin{eqnarray}\label{hylcft}
{\bf H}_\text{YLCFT} = {\bf L}_0 + {\bar{\bf L}}_0  - \frac{c}{12}\,,
\end{eqnarray}
while the spatial ($\tx$-direction in Fig.\ref{cylinder}) momentum is ${\bf P}_\tx = {\bf L}_0-{\bar{\bf L}}_0$. In what follows we limit attention to the states with zero spatial momentum ${\bf P}_\tx =0$. In this sector the eigenvalues of the YLCFT Hamiltonian \eqref{hylcft} are of the form
\begin{eqnarray}\label{encft}
-\frac{c}{12} + 2\Delta_\phi + 2l \qquad \text{and} \qquad
-\frac{c}{12} + 2l
\end{eqnarray}
for the descendants of the level $l=0,1,2,3,...$ of $ | \phi\rangle$ and $ | I \rangle$, respectively; here again $c=c_\text{YL}=-\frac{22}{5}$ and $\Delta_\phi=-\frac{1}{5}$.

\subsection*{Energy levels at large  $R$}

We assume that the eigenstates $| n \rangle_R,\ n=0,1,2,3,\dots $ of the finite-size Hamiltonian \eqref{hift} are labeled in the order of increasing eigenvalues $E_n(R)$, so that $ |0 \rangle_R$ is the ground state, $ | 1 \rangle_R$ is the first excited state, etc.  At pure imaginary $\xi$ the eigenvalues are either real or appear in complex conjugate pairs, so in fact we order $E_n(R)$ according to their real parts (with arbitrary attribution when the real parts are equal), $\Re e E_n(R) \leq \Re e E_{n+1}(R)$.
As is convenient for the massless flows, we introduce the functions
\begin{eqnarray}\label{cndef}
C_n(R) = - \frac{6R}{\pi}\,( E_n(R)-f_0 \,R )
\end{eqnarray}
which approach constants in both UV and IR limits, $R\to 0$ and $R\to \infty$, respectively. For the ground state level $E_0(R)$ the limiting values $C_0(R\to 0)$ and $C_0(R\to\infty)$ coincide with the "effective central charges" $c_\text{eff}$ of the UV fixed point ${\cal A}_\text{ICFT}$ and IR fixed point ${\cal A}_\text{YLCFT}$, respectively. We loosely refer to the functions \eqref{cndef} as the effective central charges associated with the levels $ | n\rangle_{(R)}$. Generally, $C_n(R)$ are expected to interpolate between their UV and IR limits,
\begin{eqnarray}
&&C_n (R)\  \to \ c^{(\text{UV})} - 24\Delta_n^{(\text{UV})}\,\qquad \text{as} \quad R\to 0\,,\\
&&C_n (R)\  \to \ c^{\,\,(\text{IR})} - 24\Delta_n^{(\text{IR})}\,\qquad \ \text{as} \quad R\to \infty\,,
\end{eqnarray}
where $c^{(\text{UV})}=c_\text{Ising}=\frac{1}{2}$ and $c^{(\text{IR})}=c_\text{YL} = -\frac{22}{5}$ are the Virasoro central charges associated with the UV and IR fixed points, while $2\Delta_n^{(\text{UV, IR})}$ are the eigenvalues of the operator ${\bf L}_0 +{\bar{\bf L}}_0$ on the state $ | n\rangle$ in the UV and IR CFT, respectively.

In these notations, the $R\to\infty$ limits of the states $| n \rangle_R$ relate to the YLCFT states as follows
\begin{eqnarray}
&&| 0 \rangle_{R} \to \ | \phi \rangle\,,\qquad | 1 \rangle_{R} \to\ | I \rangle\,,\qquad
| 2 \rangle_{R}\ \to\ N_2\,{\bf L}_{-1}{\bar{\bf L}}_{-1}| \phi \rangle\,,\nonumber\\
&&| 3 \rangle_{R} \to N_3\, {\bf L}_{-1}^2{\bar{\bf L}}_{-1}^2| \phi \rangle\,,\qquad | 4 \rangle_{R} \to \ \ N_4\,{\bf L}_{-2}{\bar{\bf L}}_{-2}| I \rangle\,,\qquad \text{etc}  \label{CFTstates}
\end{eqnarray}
with the coefficients $N_n$ (e.g. $N_2 = \frac{5}{2}$, $N_3=\frac{25}{12}$, $N_4=\frac{5}{11}$, etc) inserted to impose the standard normalization $\langle \; n |n'\rangle=\delta_{n,n'}$. The limiting values of the functions $-C_n:=-C_n(R=\infty)$ are
\begin{eqnarray}\label{cn}
&&-C_0 = -\frac{2}{5}=-0.4\,, \quad  -C_1=\frac{22}{5} =4.4\,,\ \  -C_2 = -\frac{2}{5}+24 = 23.6\,,\nonumber\\
&&-C_3 = -\frac{2}{5}+48 = 47.6\,,\quad -C_4=\frac{22}{5}+48 = 52.4\,,   \label{cnir}
\end{eqnarray}
The rate of approach of $C_n(R)$ to these limiting values is controlled by the effective action \eqref{aeff0}, as we discuss below.

\subsection*{$C_n(R)$ from TFFSA}

\begin{figure}[!h]
\caption{5 lowest energy levels at $\xi^2= - 0.035846$, this value is very close to the critical point $-\xi_0^2$, where the theory is gapless.}
\centering
\includegraphics[width=0.8\textwidth]{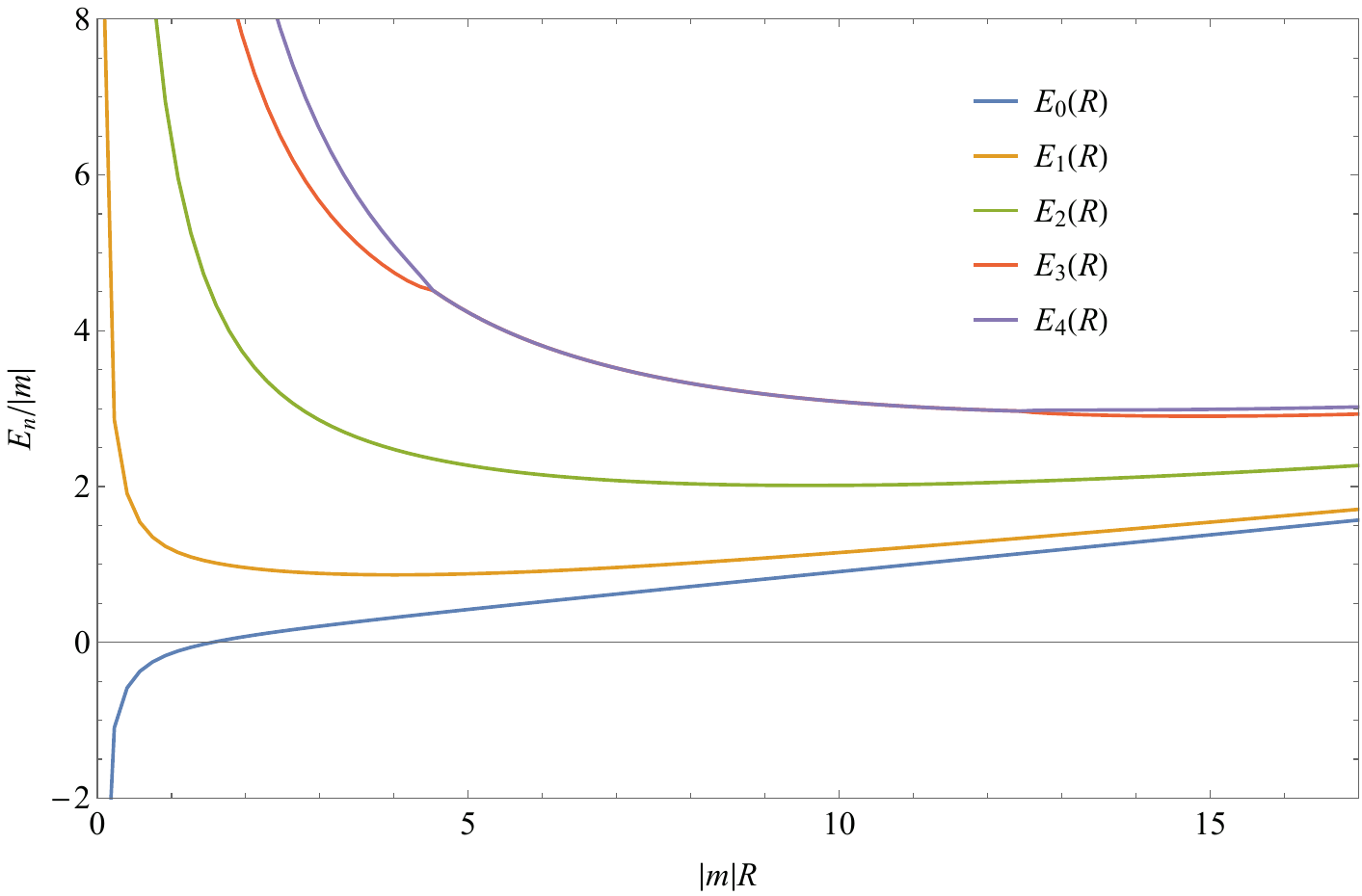}
\label{SpectrumCritical}
\end{figure}

The first five energy levels $E_n(R)$ of \eqref{ift} at $\xi^2=-0.035846$, which we believe to be very close to exact YL point $-\xi_0^2$ (see Sec.4), are shown in Fig.\ref{SpectrumCritical}. The data was obtained numerically, using TFFSA (see Sec.1) with
the truncation level $L=13$.
Note that $E_0(R)$, $E_1(R)$ and $E_2(R)$ are real for all values of $R$, whereas $E_3(R)$ and $E_4(R)$ turn into a complex conjugated pair at some intermediate values of $R$ (Plots in Fig.\ref{SpectrumCritical} show the real parts). This phenomenon is typical to the higher levels: while
taking real values for sufficiently large as well as at sufficiently small $R$, the energies $E_n(R)$ with $n>4$ form a complicated pattern of complex conjugated pairs at intermediate $R$. Whereas full understanding of pattern remains an interesting open problem, below we present partial explanation of the intricate interplay of the levels $E_3$ and $E_4$ in Fig.\ref{SpectrumCritical} (see the last subsection of this Section).

\begin{figure}[!h]
\caption{Comparison of truncation effects for different energy levels, the plots show the deviations $\Delta E_n(R) = E^{(L=13)}_n(R)-E^{(L=12)}_n(R)$, of data at different truncation levels. The truncation effect are more prominent for higher energy levels.}
\centering
\includegraphics[width=0.8\textwidth]{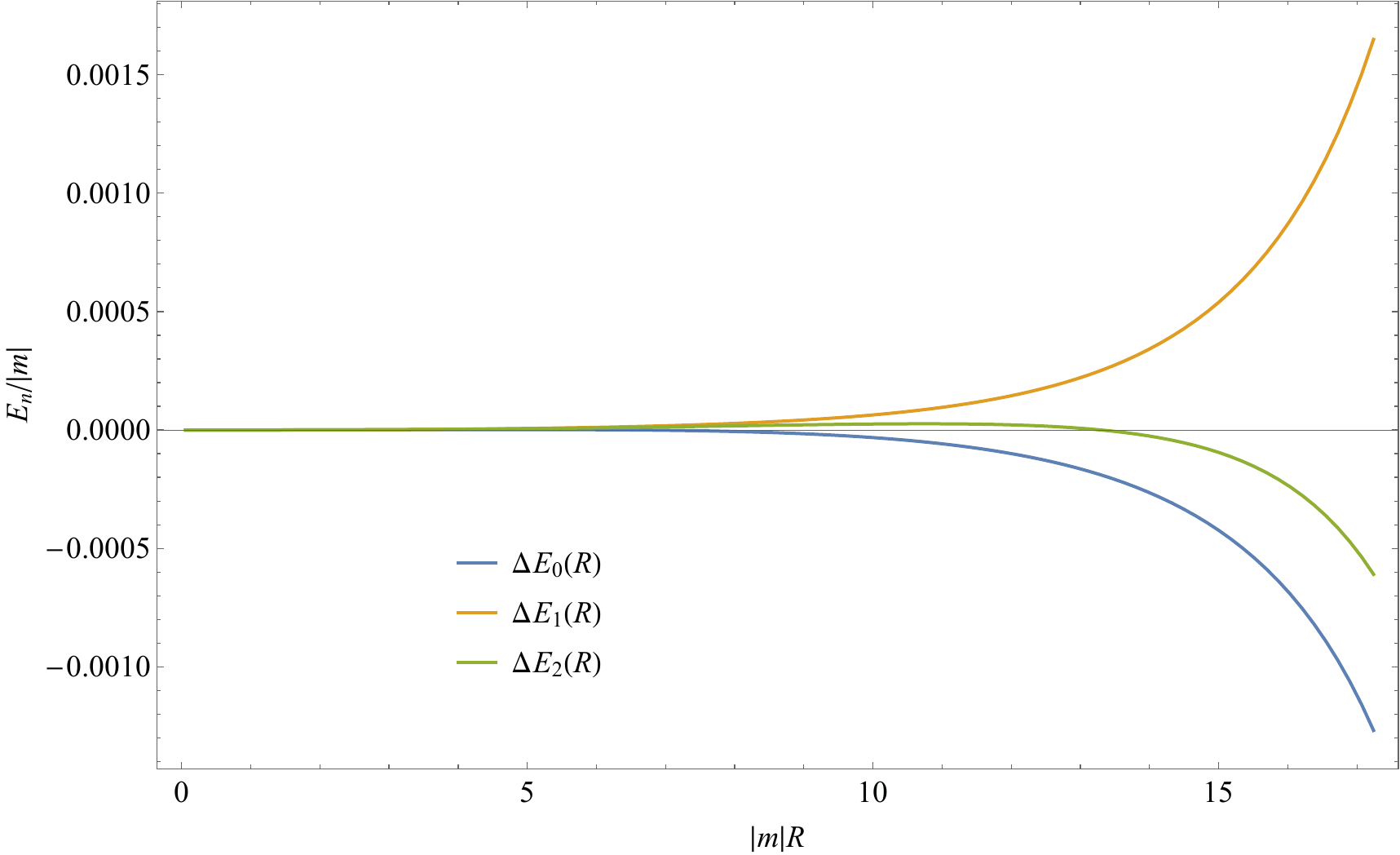}
\label{TruncationEffectCritical}
\end{figure}

\begin{figure}[!h]
\caption{Plots of the $-C_n(R)$ at $\xi^2= - 0.035846$, the dashed lines show $-C_n$ from the infrared CFT. Note that $-C_n(R)$ grows with energy levels.}
\centering
\includegraphics[width=0.8\textwidth]{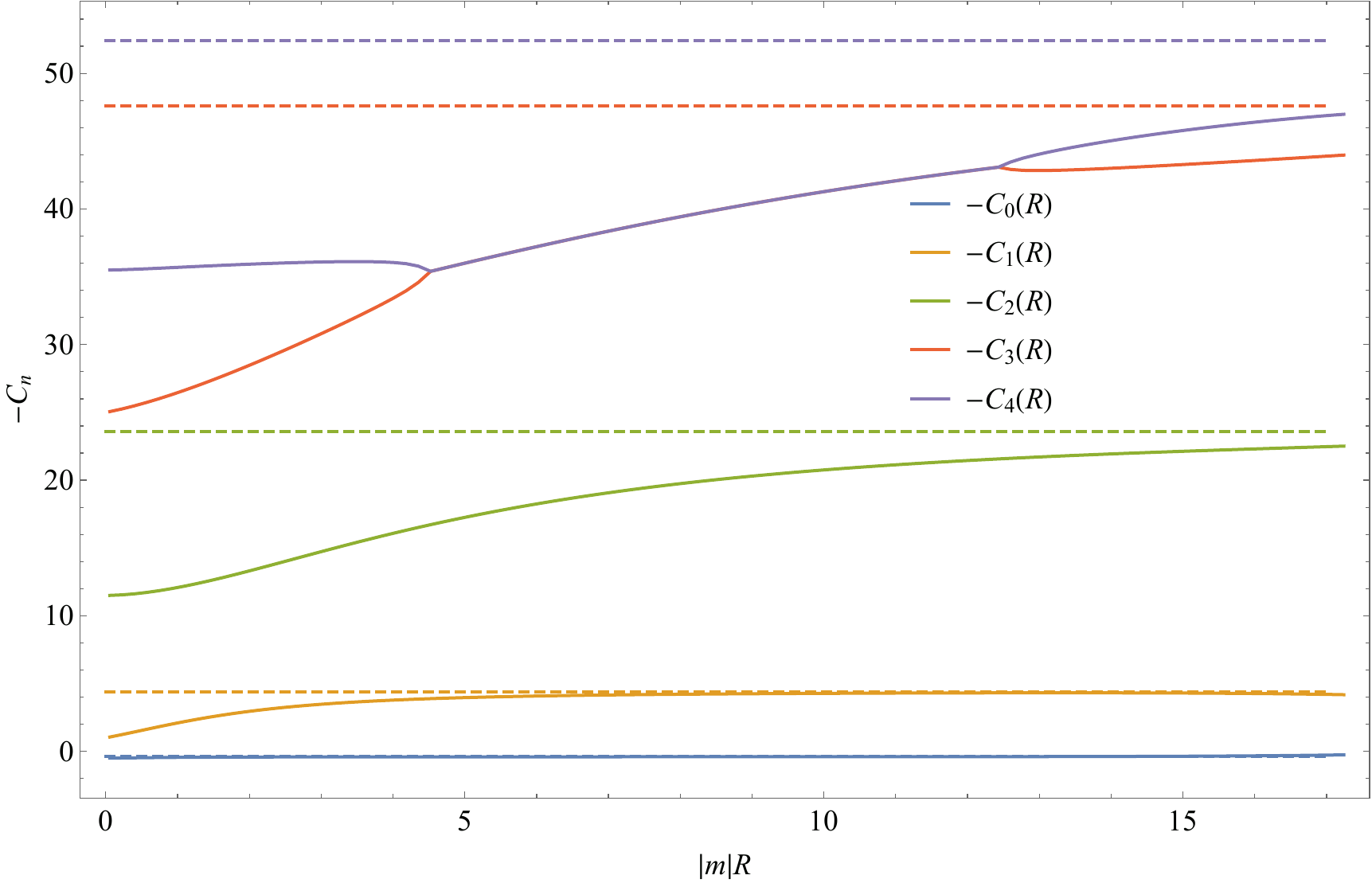}
\label{CfunctionCritical}
\end{figure}

In Fig.\ref{SpectrumCritical} we limit attention to the interval $R=[0:17.5]$ because at larger $R$ the quality of the TFFSA data rapidly deteriorates (see Fig.\ref{TruncationEffectCritical}).
Nonetheless, even in this interval the ground state energy $E_0(R)$ clearly develops linear asymptotic $E_0(R)\to f_0 R$, with the slope
\begin{eqnarray}\label{f0estimate}
f_0 \approx 0.092746...\,,
\end{eqnarray}
in agreement with the result of \cite{fonseca2003ising}. This gives the estimate of the "cosmological" parameter $f$ in \eqref{aeff0} at the YL point, $f_0:=f(-\xi_0^2)$. The behavior of the higher levels $n=2,3,4$ is consistent with
the expected large-$R$ YLCFT form, Eq.\eqref{hircft}. This is seen better in Fig.\ref{CfunctionCritical}, where the associated functions $-C_n(R)$ are plotted.
The functions $-C_n(R)$ seem to approach the expected large-$R$ limits \eqref{cnir}, although the deviations are greater for the higher levels. We attribute these deviations to the contributions of the irrelevant terms in the effective action \eqref{aeff0} at the YL critical point.

\subsection*{Effective action and large  $R$ expansion}

The irrelevant terms in the effective action \eqref{aeff0} generate large-$R$ expansions of the energy levels
$E_n(R)$. In the leading order in the couplings $a_i$ we have
\begin{eqnarray}\label{enexp}
{\cal E}_n(R):=E_n(R)-f_0 R =- \frac{\pi\,C_n}{6 R} + \sum_i \, 2\pi a_i\,[O_i]_{nn}\left(\frac{2\pi}{R}\right)^{2\Delta_i-1} + \dots \,,
\end{eqnarray}
where $C_n$ are the coefficients \eqref{cn}, and the numerical coefficients $[O_i]_{nn}$ are related to the diagonal matrix elements
\begin{eqnarray}\label{onn}
 [O_i]_{nn}:=\langle  n| O_i | n \rangle_\text{CFT}\,\left(\frac{R}{2\pi}\right)^{2\Delta_i}
\end{eqnarray}
in the YLCFT. The dots in \eqref{enexp} stand for the higher-order contributions.

The dominating correction in \eqref{enexp} clearly comes from the operator $T{\bar T}$ in \eqref{aeff1}. In fact, the higher orders in $\alpha$, up to the order $\alpha^4$, can be explicitly taken into account using the $T{\bar T}$ deformation formula \eqref{ealpha}.
This is done as follows. Consider first the fictitious effective action \eqref{aeff1} with $\alpha=0$. In such theory the leading correction correction in \eqref{enexp} would be determined by the operator $\Xi$, i.e.
\begin{eqnarray}\label{betaterm}
{\cal E}_n^{(0)}(R):={\cal E}_n^{(\alpha=0)}(R) = - \frac{\pi\,C_n}{6 R} + \beta_0\,\Xi_{nn}\left(\frac{2\pi}{R}\right)^\frac{33}{5} + O(R^{-\frac{53}{5}})
\end{eqnarray}
or
\begin{eqnarray}\label{cn0}
C_n^{(0)} = C_n - 12 \beta_0\,\Xi_{nn} \,\left(\frac{2\pi}{R}\right)^\frac{28}{5} + O(R^{-\frac{48}{5}})\,,
\end{eqnarray}
where $\Xi_{nn}$ is the diagonal matrix element \eqref{onn} of the operator $\Xi$ in the YLCFT, and $\beta_0$ is the value of $\beta(\xi^2)$ for massless flow. Matrix elements $\langle n| \Xi | n \rangle$ in the YLCFT, for the first few $n$, are computed in Appendix \ref{AppendixCPT},
\begin{eqnarray}\label{Xinn}
\Xi_{00} = \frac{1}{1200^2}\,\mathbb{C}_{\phi\phi}^\phi\,,  \ \  \Xi_{22}=\left(\frac{601}{2^3 \cdot 3 \cdot 5^3}\right)^2 \mathbb{C}_{\phi\phi}^{\phi}, \ \  \Xi_{33}=\left(\frac{56417}{2^4 \cdot 3 \cdot 5^4}\right)^2 \mathbb{C}_{\phi\phi}^\phi\,, \ \
\end{eqnarray}
where $\mathbb{C}_{\phi\phi}^\phi$ is the constant \eqref{Cphi}, while $\Xi_{11}=\Xi_{44}=0$.

Now, the contributions of the $T\bar T$ term in \eqref{aeff1} can be taken into account (up to the order $\alpha^4$) via the TTbar deformation formula \eqref{ealpha}, which amounts to replacing $R$ in \eqref{betaterm} by
\begin{eqnarray}
{\tilde R} = R -\alpha {\cal E}(R)
\end{eqnarray}
i.e.
\begin{eqnarray}\label{enalphabeta}
{\cal E}_n(R) = - \frac{\pi\,C_n}{6 (R-\alpha \,{\cal E}_n(R))} + \beta_0\,\Xi_{nn}\,\left(\frac{2\pi}{R-\alpha\, {\cal E}_n(R)}\right)^\frac{33}{5} + O(R^{-\frac{43}{5}})\,.
\end{eqnarray}
This formula implicitly defines ${\cal E}_n(R)$ as a series in inverse fractional powers of $R$, which faithfully reproduces the
energy levels of the effective theory \eqref{aeff1} up to the order indicated in \eqref{enalphabeta}. It can be compared to the
TFFSA data represented in Fig.\ref{CfunctionTTbar} and Fig.\ref{beta_fitting} to estimate the coupling parameters $\alpha$ and $\beta$.

\subsection*{Estimating $\alpha$ and $\beta$}

\begin{figure}[!h]
\caption{Plots of TFFSA data for $C_0(R)$ (in red line), and the plots of \eqref{CnTTbarUndressing} against $\tilde R$ (TTbar-undressed $C_0^{(0)}(R)$), with few sample values of $\alpha_0$.}
\centering
\includegraphics[width=0.75\textwidth]{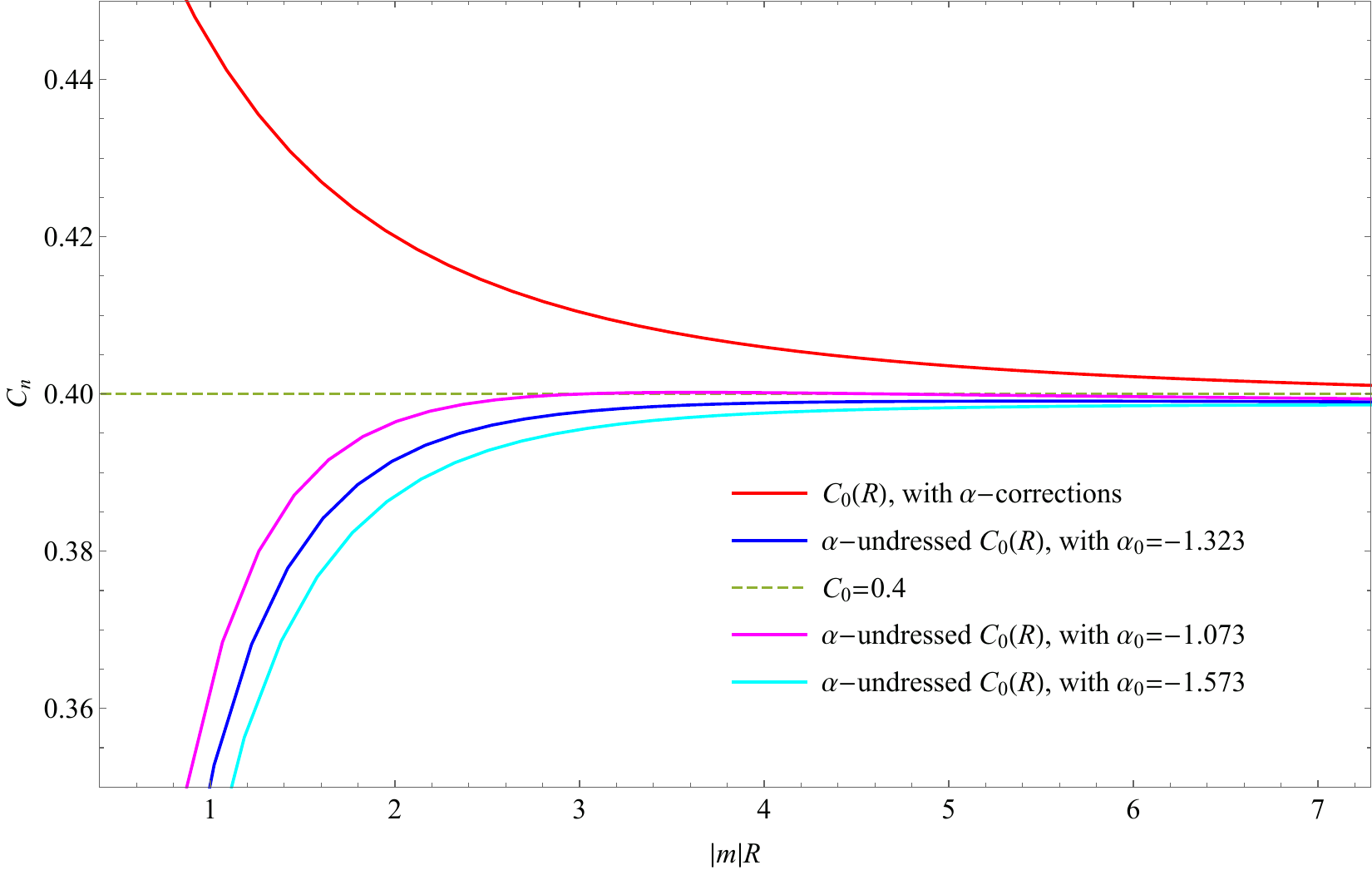}
\label{C0functionTTbar}
\end{figure}

\begin{figure}[!h]
\caption{Plots of TFFSA data for $C_1(R)$ (in red line), and the plots of \eqref{CnTTbarUndressing} against $\tilde R$ (TTbar-undressed $C_1^{(0)}(R)$), with few sample values of $\alpha_0$. The estimating value of $\alpha_0$ can be read from the blue line.}
\centering
\includegraphics[width=0.75\textwidth]{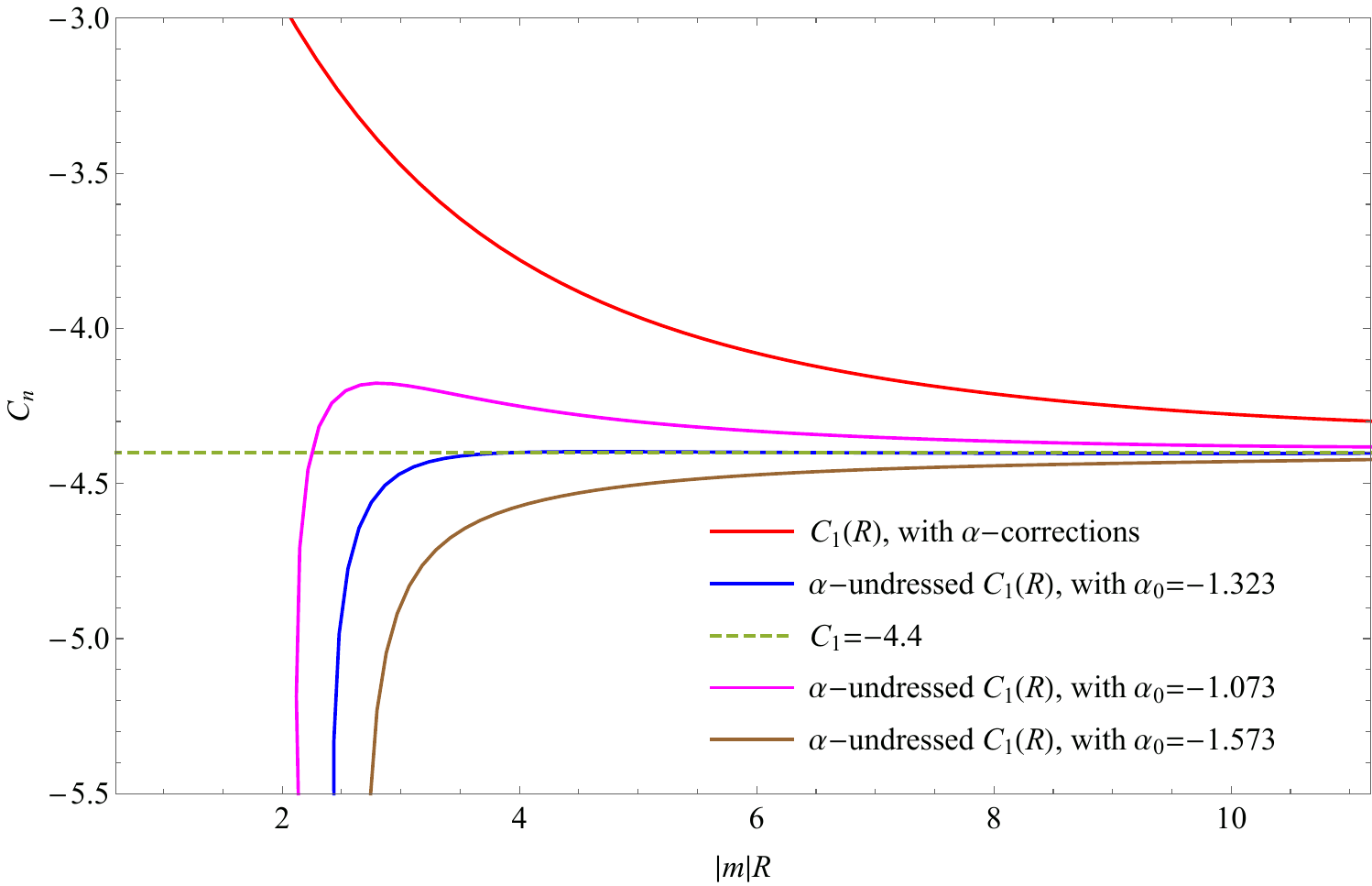}
\label{CfunctionTTbar}
\end{figure}

\begin{figure}[!h]
\caption{Plots of TFFSA data for $C_2(R)$ (in red line), and the plots of \eqref{CnTTbarUndressing} against $\tilde R$ (TTbar-undressed $C_2^{(0)}(R)$), with few sample values of $\alpha_0$.}
\centering
\includegraphics[width=0.75\textwidth]{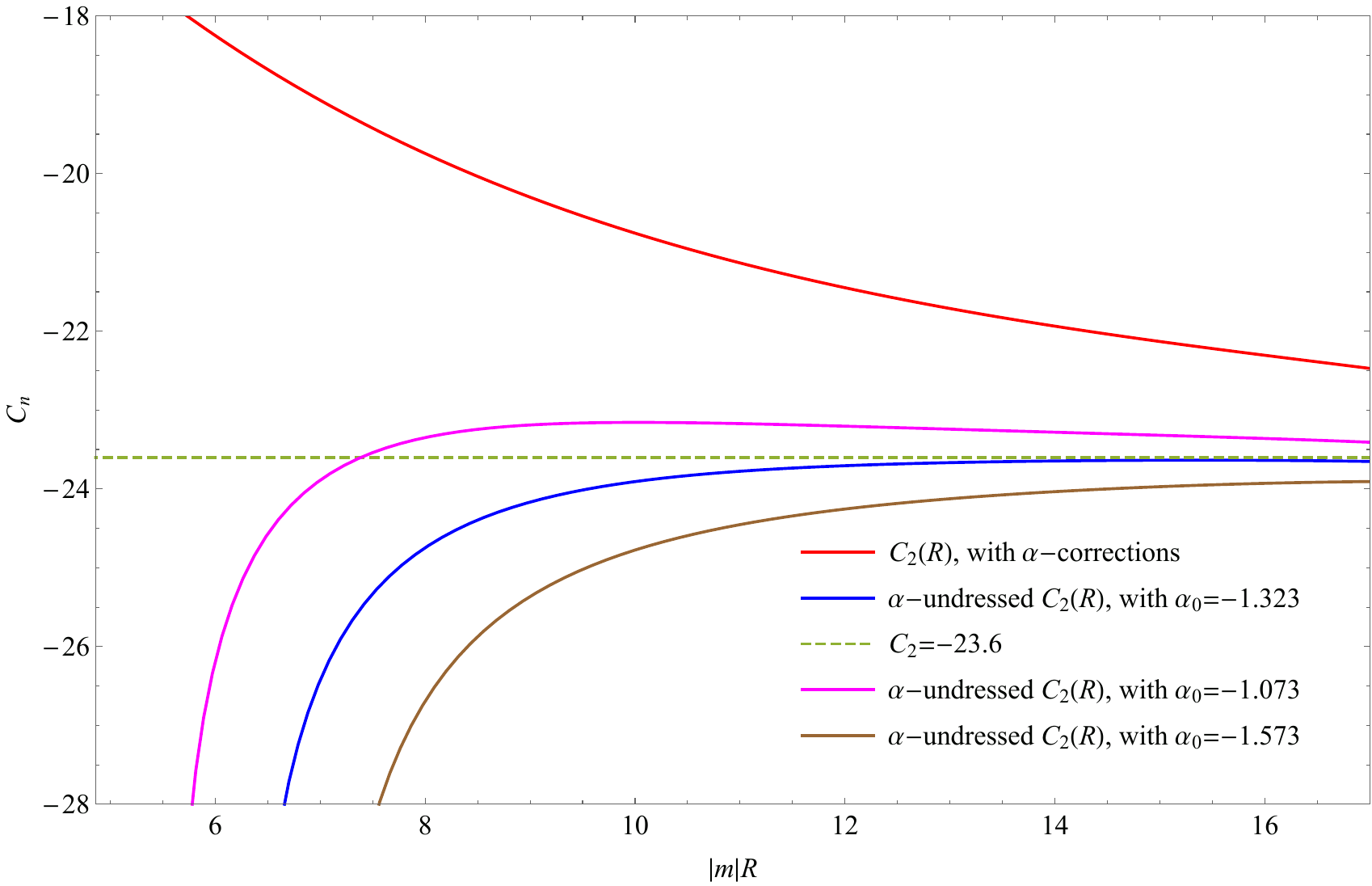}
\label{C2functionTTbar}
\end{figure}

Eq.\eqref{enalphabeta} can be used as the fitting formula for the energy levels ${\cal E}_n(R)=E_n(R)-f_0 R$ of the IFT obtained by TFFSA, to determine the coupling parameters. We found slightly different approach to be advantageous. Given the TFFSA data for the energy levels ${\cal E}_n(R)$ we plot
\begin{eqnarray}\label{CnTTbarUndressing}
C_n^{(0)} = - \frac{6}{\pi}\,(R-\alpha{\cal E}_n(R))\,{\cal E}_n(R)
\end{eqnarray}
against ${\tilde R}=R-{\cal E}_n(R)$. Under suitable choice of $\alpha$ these plots should reproduce the "$\alpha$-undressed"
levels $C_n^{(0)}(R)$ defined via Eq.\eqref{cn0}, which at large $R$ converge very rapidly to the limiting constants \eqref{cn}, with the leading deviation determined by the $\beta$-term in \eqref{cn0}. These plots are given in Fig.\ref{C0functionTTbar}, Fig.\ref{CfunctionTTbar},
Fig.\ref{C2functionTTbar} for the first three levels, with few test values of $\alpha$ (The figures also show the corresponding functions $C_n(R)$, Eq.\eqref{cndef} from the data).

We observe that with $\alpha$ close to $-1.3$ the deviation of $C_n^{(0)}(R)$ from $C_n$ at large $R$ is indeed very small. Moreover, we note that for the first excited state coefficient $\Xi_{11}$ vanishes. That is, with a correct choice of $\alpha$ the approach of $C_1^{(0)}(R)$ to $-4.4$ can be made even faster\footnote{At least as fast as $O(R^{-10})$, the degree at which the contribution of operator $L_{-6}\bar L_{-6} I$ would interfere with $\alpha^5$ term generated by $T \bar T$ flow.}. For this reason we take the value of $\alpha$ minimizing the deviation $C_{1}^{(0)}(R)$ from $-4.4$ at larger $R$ as the best estimate of $\alpha$. This way we find
\begin{eqnarray}\label{alphaestimate}
\alpha |m|^2 = -1.32(5)\,,
\end{eqnarray}
where we restored the scale parameter $m$ from \eqref{ift} (previously set to 1) to emphasize the units, and the uncertainty figure
reflects the spread of the values which give the fastest decay of $C_1^{(0)}(R)+4.4$.

\begin{figure}[!h]
\caption{Fitting of $\beta$-contribution to $C^{(0)}_2(\tilde R)$ with the power law $\tilde R^{-\frac{28}{5}}$. Note that now the horizontal axis is $\tilde R = R -\alpha_0 \mathcal E_2 $, and the best fit yields $\beta_0$ in \eqref{aeff1}.}
\centering
\includegraphics[width=0.75\textwidth]{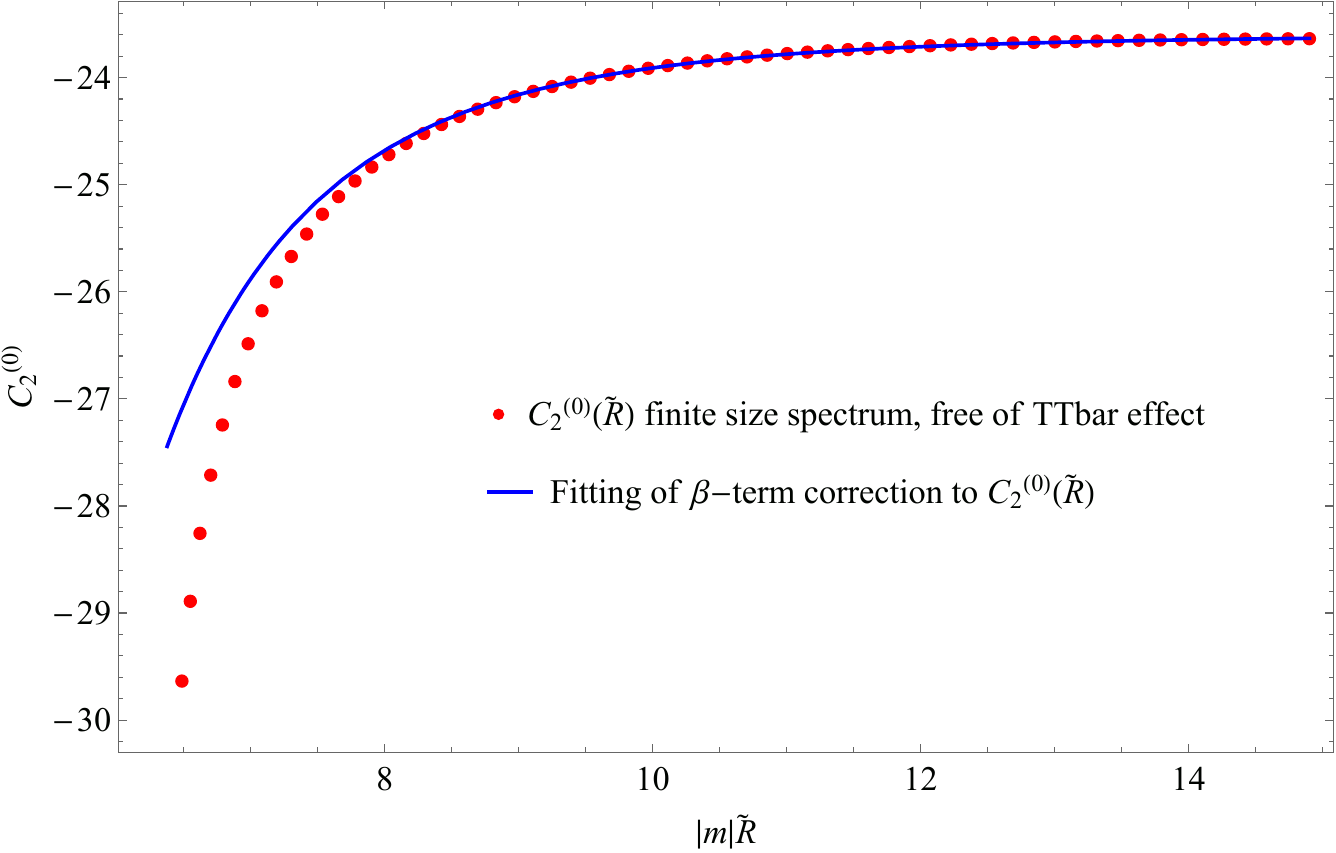}
\label{beta_fitting}
\end{figure}

With this estimate of $\alpha$ one can use the large-$R$ behavior of $C_0^{(0)}(R)$ and $C_2^{(0)}(R)$ to estimate $\beta$. We note that the coefficient $\Xi_{00}$ is numerically very small as compared to $\Xi_{22}$ ($\Xi_{22}/\Xi_{00} \approx 50,000$). Therefore, at large $R$ the deviation $C_0^{(0)}(R)-C_0$ is expected to be much smaller than $C_2^{(0)}(R)-C_2$. Indeed, Fig.\ref{CfunctionTTbar} shows that with $\alpha=-1.32$ the deviation of the plot of $C_0^{(0)}(R)$ from $0.4$ becomes exceedingly small at $R$ greater than $5$, while at smaller $R$ the contribution of the $\beta$-term may become comparable to the the higher order terms neglected in \eqref{betaterm}. On the other hand, the deviation of $C_2^{(0)}(R)$ from $-23.4$ remains quite appreciable even at $R \geq 8$, and we can compare it with the $\beta$-term in \eqref{betaterm}. Fig.\ref{beta_fitting} shows the best fit
of the Eq.\eqref{betaterm} to $C_2^{(0)}(R)$ obtained from data, in the interval $R=[8 : 12]$, with $\beta$ taken as the fitting parameter. The result is
\begin{eqnarray}\label{betaestimate}
\frac{\beta_0}{2\pi}\, |m|^{\frac{28}{5}} = +0.72 \pm 0.06\,,
\end{eqnarray}
where again we made units explicit, and the uncertainty reflects the dependence on the change of the fitting interval.

While Eq.\eqref{betaestimate} is the first estimate of the coupling $\beta$ in the effective action \eqref{aeff1} at
the YL critical point $\xi^2=-\xi_0^2$, the parameter $\alpha$ was previously estimated in Ref.\cite{fonseca2003ising}.
That work presents two independent evaluations of $\alpha_0$. One is based on the analysis of the finite size ground state
energy, $\alpha_0=-1.1(1)$ (Eq.(7.4) of \cite{fonseca2003ising}), and the other from the singular part of the vacuum energy
density near the critical point, $\alpha_0=-1.2(2)$ (Eq.(7.3) of \cite{fonseca2003ising})\footnote{We adjust the results of
\cite{fonseca2003ising} to our notations here: Our $\alpha$ in \eqref{aeff1} differs from that in \cite{fonseca2003ising}
by the factor of 4.}. Our result \eqref{alphaestimate} agrees with the second of these figures within the stated accuracy, but slightly disagrees with
the first one. We believe that our result here is more reliable. While the first estimate in \cite{fonseca2003ising}
was made by fitting the ground state energy, while \eqref{alphaestimate} here was obtained from the first excited state, where
the dominating contribution of the $\alpha$-term ($\sim \alpha C_n^2/R^2$) is $\approx 100$ times greater.

\subsection*{Levels 3 and 4. Level crossing via the $\beta$-term}

\begin{figure}[htb]
\centering
\includegraphics[width=0.8\textwidth]{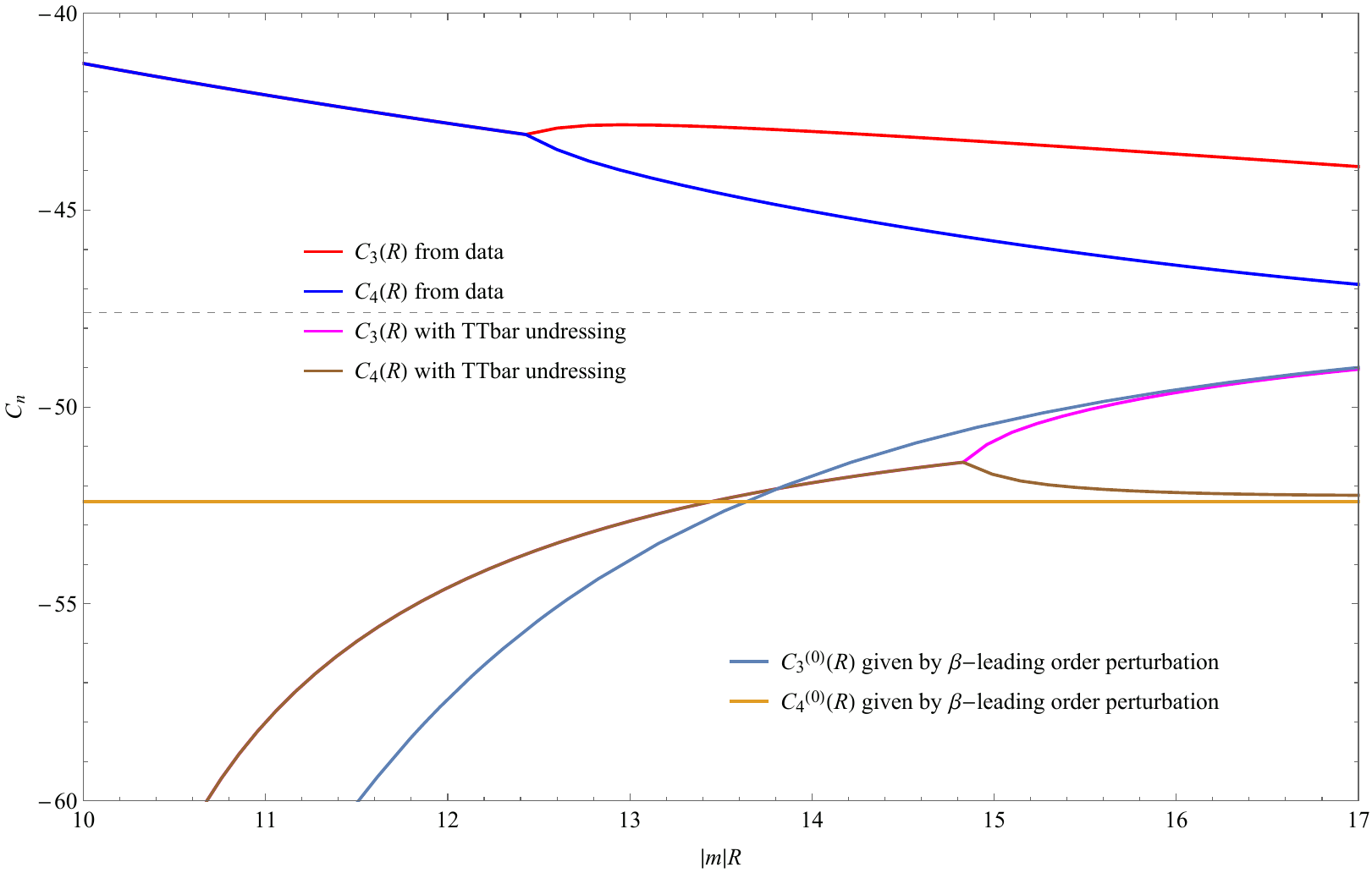}
\caption{"Level crossing" of $C_3(R)$ and $C_4(R)$. TFFSA data for $C_3(R)$ and $C_4(R)$ are shown in red and blue, respectively. The real parts are shown below the crossing point at $R \approx 12.4$, where the levels turn into a complex conjugate pair. The dotted lines show the constant $C_3$ and $C_4$ from Eq.\eqref{cn}. The "TTbar-undressed" functions $C^{(0)}_3(R)$ and $C^{(0)}_4(R)$ are shown in magenta and brown respectively. Light blue and light brown represent approximation \eqref{cn0} for $C^{(0)}_3(R)$ and $C^{(0)}_4(R)$.} \label{Fig.C34}
\end{figure}

While the energies $E_0(R)$, $E_1(R)$, $E_2(R)$ remain real at all $R$, the next two levels exhibit more intricate behavior, see Fig.\ref{SpectrumCritical}. Although the eigenvalues $E_3(R)$ and $E_4(R)$ are real at small as well as at large $R$ (as is demanded by the UV and IR CFT limits of the RG flow) in the crossover region $  4 \lessapprox R \lessapprox 13$ they turn into a complex conjugate pair. The corresponding functions $C_3(R)$ and $C_4(R)$ are shown in Fig.\ref{Fig.C34}.
\begin{figure}[htb]
\centering
\includegraphics[width=0.8\textwidth]{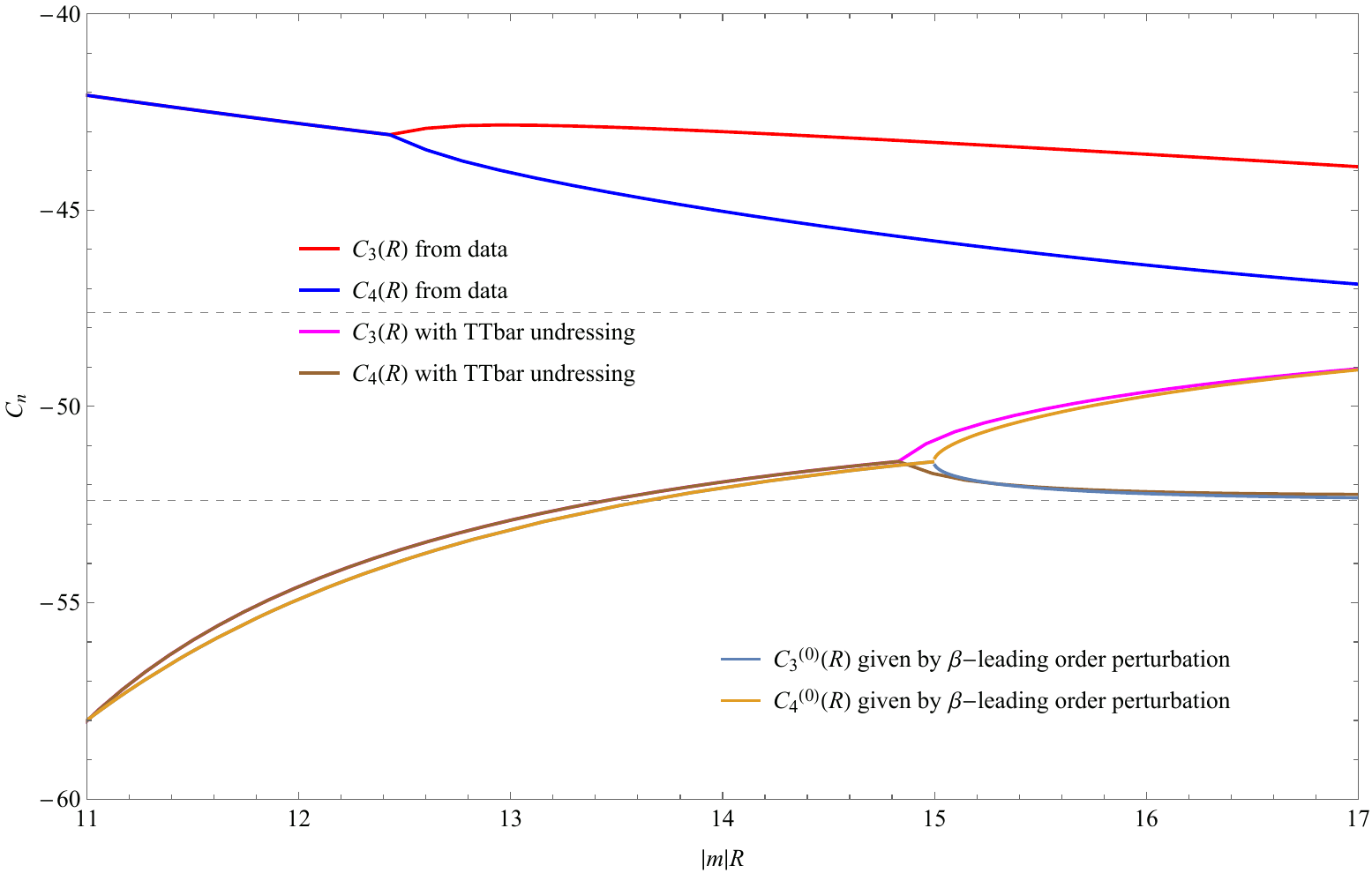}
\caption{Same as in Fig.\ref{Fig.C34}, but now light blue and light brown show eigenvalues of the matrix \eqref{c34matrix}.} \label{Fig.C34c}
\end{figure}
As an integrable theory is unlikely to develop such behavior\footnote{See however \cite{camilo2021factorizable}}, it is tempting to attribute the large $R$ part of this pattern to the effect of the operator $\Xi$ in \eqref{aeff1}, which is the lowest dimension operator breaking integrability. At large $R$, $C_3(R)$ and $C_4(R)$ are expected to approach constants $C_3=-47.6$ and $C_4=-52.4$, respectively (see Eq.\eqref{cn}). Although the cutoff bound $R<17$ (beyond which the TFFSA data become less reliable) does not allow to see this expected $R\to\infty$ asymptotic, the behavior of $C_3(R)$ and $C_4(R)$ in Fig.\ref{Fig.C34} is at least consistent with it. The functions $C_3(R)$ and $C_4(R)$ remain real (with $-C_4(R)>-C_3(R)$) at all $R$ above the "level crossing" point $R_{34}\approx 12.4$, where the levels collide and become a complex conjugate pair at $R<R_{34}$. (Here we ignore another level crossing at much smaller $R_{34}' \approx 4.5$, which hardly can be explained in terms of the perturbative analysis bases of the IR effective action \eqref{aeff0}.) Qualitatively, this behavior of $C_3(R)$ and $C_4(R)$ agrees with what one expects from the contribution of the operator $\Xi$ in \eqref{aeff1}.

Consider "$\alpha$-undressed" functions $C_3^{(0)}(R)$ and $C_4^{(0)}(R)$ which correspond to the energy levels of \eqref{aeff1} with $\alpha=0$. Their large-$R$ behavior is expected to follow \eqref{cn0}. The separation between the asymptotic values $C_3$ and $C_4$
is relatively small, and with the matrix elements $\Xi_{33}$ and $\Xi_{44}$ from \eqref{Xinn} and positive $\beta$ the values
of $C_3(R)$ and $C_4(R)$ get yet closer when one goes from large to smaller $R$. Eventually, at some $R$ the separation become very small. In this domain of $R$ the formulae \eqref{cn0} (which are based on the first order perturbation in $\beta$) do not apply. Instead, one has to use the perturbation theory for near-degenerate levels, which involve the off-diagonal matrix elements $\Xi_{34}
=\Xi_{43}$. Simple analysis of the corresponding secular equation shows that the levels $C_3(R)$ and $C_4(R)$ would collide
at some $R$, and turn into complex conjugate pair at lower $R$. On the qualitative level, this nicely agrees with the
level-crossing patters in Fig.\ref{Fig.C34}. Moreover, with the our previous estimates of $\alpha_0$ and $\beta_0$ (Eq's.\eqref{alphaestimate} and \eqref{betaestimate}), TFFSA data for $C_3(R)$ and $C_4(R)$ are reproduced very closely.

In Fig.\ref{Fig.C34} we show, along with the direct TFFSA data for $C_3(R)$ and $C_4(R)$, the "$\alpha$-undressed" functions $C^{(0)}_3(R)$ and $C_4^{(0)}(R)$ obtained from the data by applying the $T\bar T$ flow formula, as explained in the previous subsection, with the estimated value of $\alpha_0$ in Eq.\eqref{alphaestimate}. Note that the approach of the "undressed" functions to the constants $C_3$ and $C_4$ looks much more convincing than that of the full $C_3(R)$ and $C_4(R)$. In the same Fig.\ref{Fig.C34}
we plot the perturbative estimate for these functions, Eq.\eqref{cn0} with the coupling $\beta$ from \eqref{betaestimate} (since $\Xi_{44}=0$ the plot of perturbative $C_4^{(0)}(R)$ from \eqref{cn0} is just the horizontal line $C_4$). One can see that the separation between these perturbative $C_3^{(0)}(R)$ and $C_4^{(0)}(R)$ becomes small at $R \approx 14$. As was already mentioned,
in this region the approximation \eqref{cn0} for $C_3^{(0)}(R)$ and $C_4^{(0)}(R)$ breaks down. Instead, in this domain the levels
$C_3^{(0)}(R)$ and $C_4^{(0)}(R)$ must be obtained by diagonalization of the $2\times 2$ matrix
\begin{eqnarray}\label{c34matrix}
\mat{C_3}{0}{0}{C_4} - 12\beta_0\,\left(\frac{2\pi}{R}\right)^\frac{28}{5}\,\mat{\Xi_{33}}{\Xi_{34}}{\Xi_{43}}{\Xi_{44}}
\end{eqnarray}
where $\Xi_{34}=\Xi_{43}$ denotes the matrix element $\langle 3 | \Xi | 4 \rangle \approx 2.37768 i$ in the YL CFT (see Appendix \ref{AppendixCPT}).
In Fig.\ref{Fig.C34c} we compare the $C_3^{(0)}(R)$, $C_4^{(0)}(R)$ obtained by "$T{\bar T}$ undressing" of the TFFSA data, with $\alpha_0$ from \eqref{alphaestimate}, as described above, to the eigenvalues of the matrix \eqref{c34matrix} with $\beta_0$ from \eqref{betaestimate}. The impressive agreement may be regarded as an independent cross-check of our estimates \eqref{alphaestimate} and \eqref{betaestimate}.

\section{Correlation Length near YL Critical Point}

Away from the YL critical point the IFT \eqref{ift} is massive. TFFSA allows one to obtain numerical results for the mass $M(\xi^2)$ of the lightest particle, which defines the correlation length $R_c=M^{-1}$. The most direct way to obtain $M$ is by analyzing the TFFSA data for the energy gap between first excited level $E_1(R)$ and ground level $E_0(R)$, which can be done for pure imaginary $h$ as well as for real values of this parameter. Below we concentrate most attention on negative $\xi^2$ between $0$ and $-\xi_0^2$, the main objective being to verify the singular expansion \eqref{Mexpansion} near the YL critical point.

\subsection*{Finite size level $E_1(R)$ and $M(\xi^2)$}

\begin{figure}[!h]
\caption{Off critical spectrum, an example at $\xi^2 = -0.0169 > -\xi_0^2$. One can observe nonvanishing gap between $E_0(R)$ and $E_1(R)$. Note that for states above $E_2(R)$ different energy eigenvalues exhibit level crossing, forming complex conjugate pairs at intermediate values of $R$ (only real parts are shown in this plot). All levels become real at sufficiently large $R$.}
\centering
\includegraphics[width=0.75\textwidth]{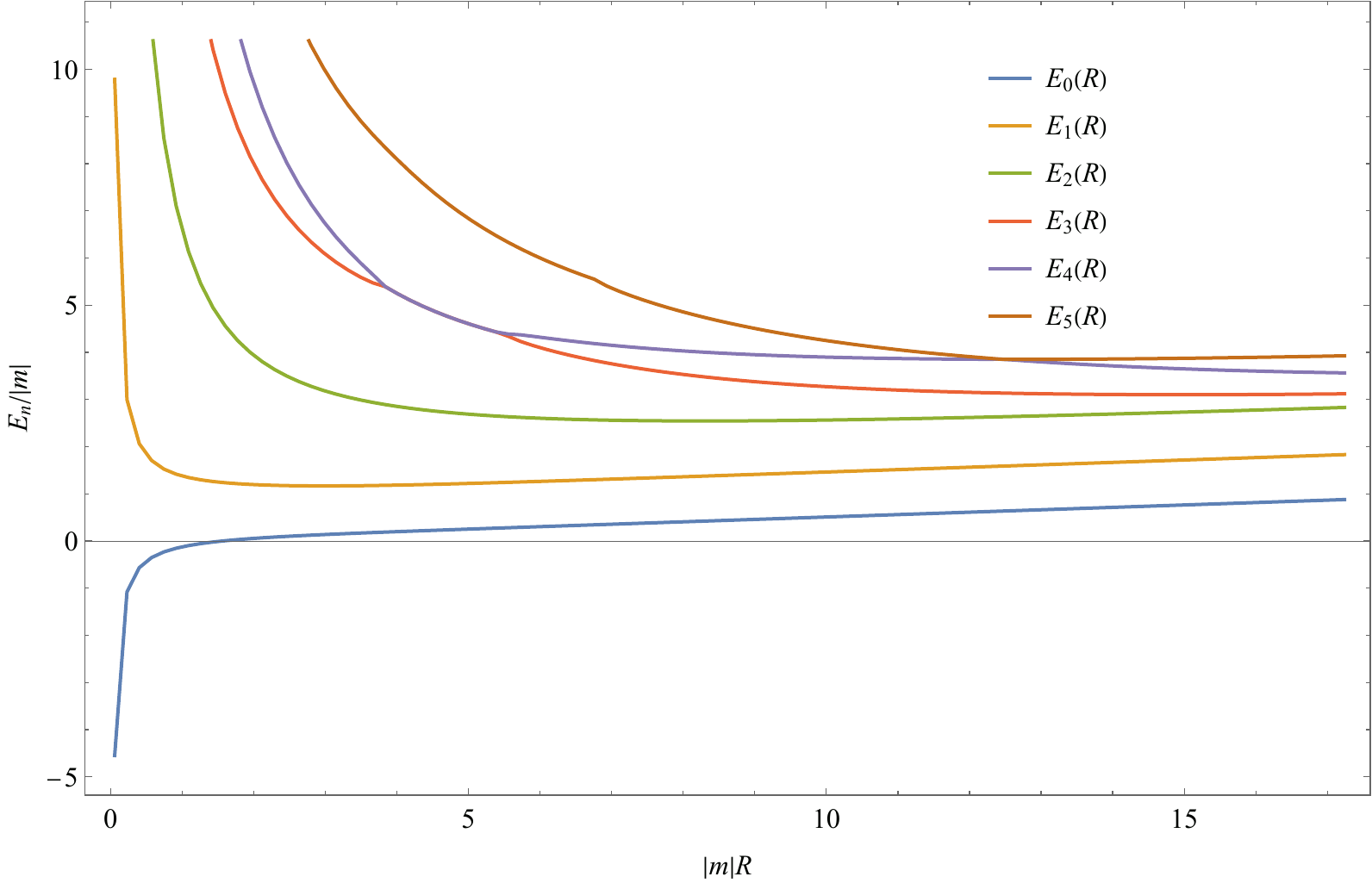}
\label{SpectrumExample}
\end{figure}

The first few levels $E_n(R)$ obtained by TFFSA with the truncation level $L=13$ (Eq.\eqref{TruncationLevelDefinition}),
at a sample value of $\xi^2$ between $-\xi_0^2$ and $0$, is shown in Fig.\ref{SpectrumExample}, where we limit attention
to the range of $|m| R < 17$, where the truncation effects remain negligible, at least for the lowest levels.
The large $R$ behavior \eqref{enir} is clearly visible, with the asymptotic form \eqref{e0ir} and \eqref{e1ir} approached
exponentially fast. In principle, the ground state $E_0(R)$ can be used to determine the vacuum energy density $F$ at a given $\xi^2$, and then \eqref{e1ir} allows one to estimate $M$. However, this straightforward approach does not produce optimal accuracy, in view of the limited range of $R$ where truncation effects are negligible. This problem becomes particularly prominent at $\xi^2$ close to the YL critical point $-\xi_0^2$.  The asymptotic decay \eqref{e0ir},\eqref{e1ir} is expected to appear at larger values of $R >> R_c$, and since $R_c(\xi^2)$ diverges near the critical point these asymptotic forms are pushed away to the domain of $R$ where the truncation effects in TFFSA become significant. Therefore, more elaborate analysis is desirable.

Much better results are obtained when taking into account the leading finite-size corrections
to \eqref{e0ir},\eqref{e1ir}. Thus, for the ground state the universal leading correction
\begin{eqnarray}
E_0(R)=F R + \frac{M}{\pi}K_1(M R)+O(e^{-2MR})
\end{eqnarray}
(see e.g. \cite{yurov1990truncated,yurov1991truncated}), where $K_1(r)$ is the Macdonald function. The above expression can be used as the fitting formula to estimate $F$ and $M$ (only one stable particle is present in the domain $-\xi_0^2<\xi^2\leq 0$). This procedure was applied in Ref.\cite{fonseca2003ising}, yielding rather accurate results (5 to 6 significant digits) for $F$ at all $\xi^2$ not too close to the critical point. However, the estimate for $M$ by this method is not sufficiently precise. In the present work we have obtained much better numerics for $M$ by analyzing the first excited level $E_1(R)$. Our procedure was as follows.

Let $S(\theta)$ be the amplitude of the elastic scattering of two lightest particles in the IFT (as before, $\theta=\theta_1-\theta_2$ denotes the rapidity difference). Then the leading finite-size corrections to {the energy gap $\Delta E = E_1 -E_0$ asymptotic (see \eqref{e0ir} and \eqref{e1ir})} can be expressed as
\cite{yurov1990truncated},\cite{Klassen:1990ub}
\begin{eqnarray}\label{SpectrumGapRemainingTerms}
\Delta E(R)= M -\frac{\sqrt{3}M}{2} \Gamma^2 e^{-\frac{\sqrt{3}}{2}M R}
 -M\int \frac{d\theta}{2\pi} \cosh\theta \big[S(\theta + \frac{\pi i}{2}) - 1 \big] e^{-M R \cosh \theta}
+ \cdots \, ,
\end{eqnarray}
where $\Gamma^2={-i}\,\text{Res}_{\theta = \frac{2\pi i}{3}}S(\theta)$ (negative at pure imaginary $\xi$) is the square of the three-particle vertex. The exhibited terms are contributions of the diagrams where one particle winds around the compactified direction once, while the dots stand for contributions with more windings which generally depend on the multi-particle scattering amplitudes. The leading term in \eqref{SpectrumGapRemainingTerms} suggests simple fitting formula
\begin{eqnarray}\label{basicfit}
\Delta E_1(R) = M + M B\,e^{-k M R}\,,
\end{eqnarray}
with $M$, $B$, and $k$ as the fitting parameters, for the TFFSA data for $\Delta E_1(R)=E_1(R)-E_0(R)$. The proximity of $k$ obtained with this fitting to $\sqrt{3}/2$ can be used for the quality control.  The fitting procedure bases on \eqref{basicfit} produces rather accurate results (in practice four to five significant digits) as long as the fitting interval of $R$ satisfies $R>>R_c$. However, it works well only when the correlation length $R_c=M^{-1}$ remains substantially smaller than the "cutoff" value $7.5$, which we imposed to keep truncation effects negligible. This holds reasonably well at $\xi^2+\xi_0^2 > 0.019$ where $R_c < 0.78$.
Closer to the critical point the correlation length becomes comparable to the cutoff and the quality of the results rapidly deteriorates. To obtain $M$ at $\xi^2$ near the critical point we used slightly refined method.

At $R \sim R_c$ the third term in the expansion \eqref{SpectrumGapRemainingTerms}, as well as the omitted higher terms, become significant. As was mentioned, these terms depend on the S-matrix of the theory. The S-matrix of IFT is generally unknown, except for special cases of $\xi^2$ where the theory is integrable. One of these cases is the close vicinity of the YL critical point. At $\xi^2$ near $-\xi_0^2$ we have $M << |m|$, and the low-energy behavior ($E\sim M$) behavior is described by YLQFT, with known factorizable S-matrix. In this limit all finite-size corrections in \eqref{SpectrumGapRemainingTerms} can be efficiently summed up via the
techniques\cite{Bazhanov:1994ft,Bazhanov:1996aq} generalizing the Thermodynamic Bethe Ansatz (TBA). {We denote $G(M R)$ the gap $E_1(R)-E_0(R)$ obtained via the generalized TBA equations (we used the results of \cite{zamolodchikov1990thermodynamic} and \cite{Bazhanov:1996aq}). The fitting formula
\begin{eqnarray}\label{E1-TBA}
\Delta E_1(R) = M + \frac{B}{3}\,G(MR)\,,
\end{eqnarray}}
with $M$ and $B$ the taken as the fitting parameters\footnote{An additional factor $B$ was introduced to mimic the $\xi^2$-dependence of the three-particle vertex $\Gamma$. For the pure YLQFT $B=3$, see Eq.\eqref{syl}.}, was used in the domain $-\xi_0^2 < \xi^2 < -0.0169$. This fitting procedure yields $M$ with reasonable accuracy of 3-4 significant digits except for the very close proximity of $-\xi_0^2$ where $R_c \sim 10$, where the accuracy falls to two digits.

\begin{figure}[htb]
\centering
\caption{Plots of $M(\xi^2)$ by fitting TFFSA data for $\xi^2 >-\xi_0^2$, with real and imaginary $h$ denoted by blue and red dots respectively. The purple line is given by mass dispersion relation (see Sec.\ref{Sec:Mdisp}).}
\includegraphics[width=0.80\textwidth]{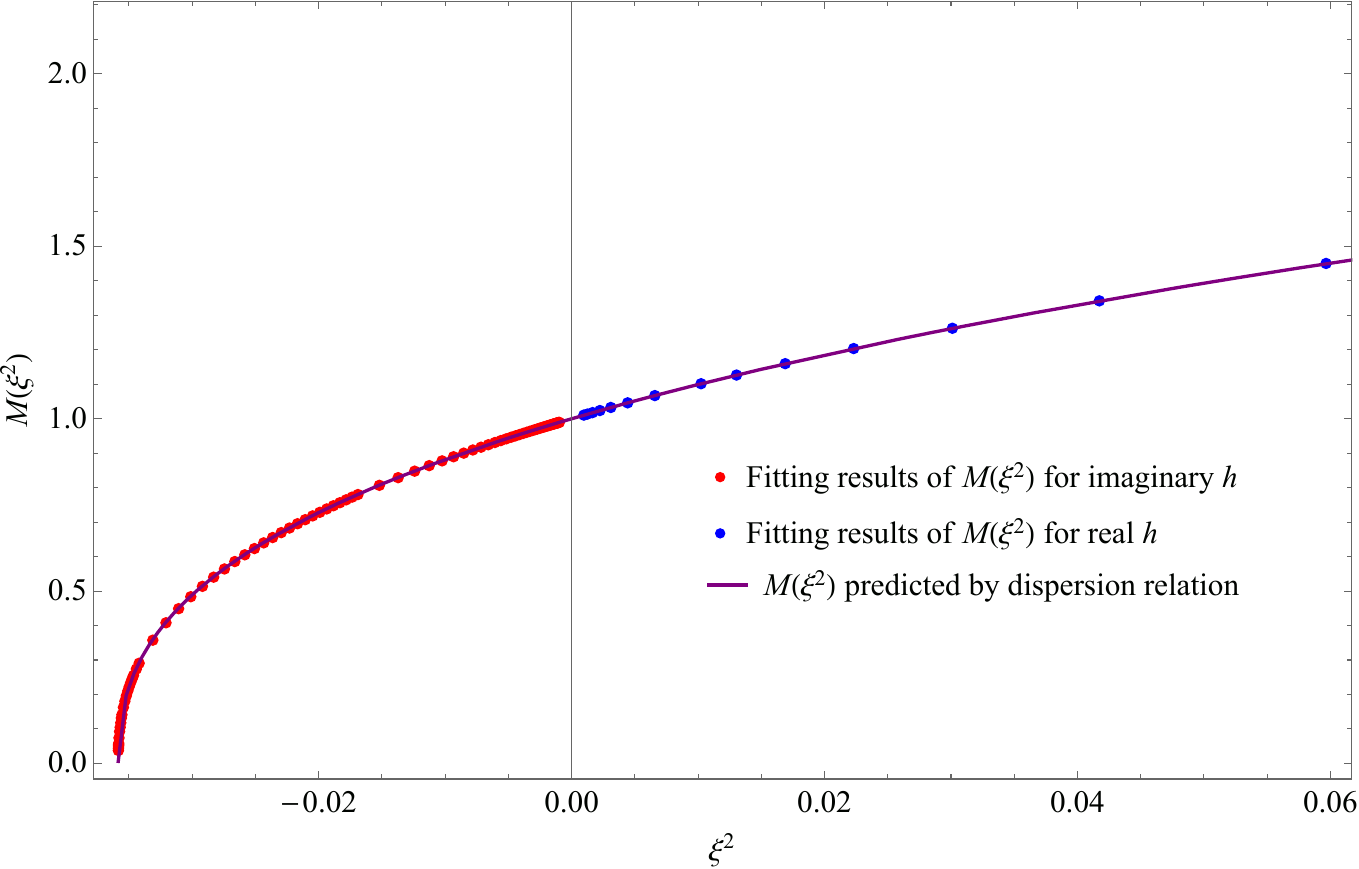}
\label{MassFittingFullxi2}
\end{figure}

Combined results for $M(\xi^2)$ obtained by these methods are presented in Fig.\ref{MassFittingFullxi2}, which also shows
some data points at positive $\xi^2$.
As expected, the points fall onto a smooth curve which agrees with known expansions around solvable points $\xi^2=0$ and $\xi^2=+\infty$.

\subsection*{Locating the Critical Point}

The location of the YL critical point was previously estimated in Ref.\cite{fonseca2003ising}, $\xi_0^2\approx 0.03583$, and independent estimate was made in \cite{BazhanovYL}, {$\xi_0^2\approx 0.03587$}. Here we use the numerical results for $M(\xi^2)$ from the previous subsection to obtain somewhat more accurate estimate.

The singular expansion \eqref{Mexpansion} suggests, in particular that the ratio $\mathcal R(\xi^2)=M(\xi^2)/(\xi^2 + \xi_0^2)^{\frac{5}{12}}$
has a finite limit at $\xi^2\to -\xi_0^2$ (with the limiting value determined by the coefficient $\lambda_1$ in \eqref{lambdaexp}, see Eq.\eqref{Mexpcoeffs}).
\begin{figure}[htb]
\caption{Data for the ratio $\mathcal R(\xi^2) =  M(\xi^2)/ (\xi^2 + \xi_0^2)^{\frac{5}{12}}$, with different values of $\xi^2_0$.}
\centering
\begin{minipage}[htb]{0.49\textwidth}
\centering
\includegraphics[width=1.0\textwidth]{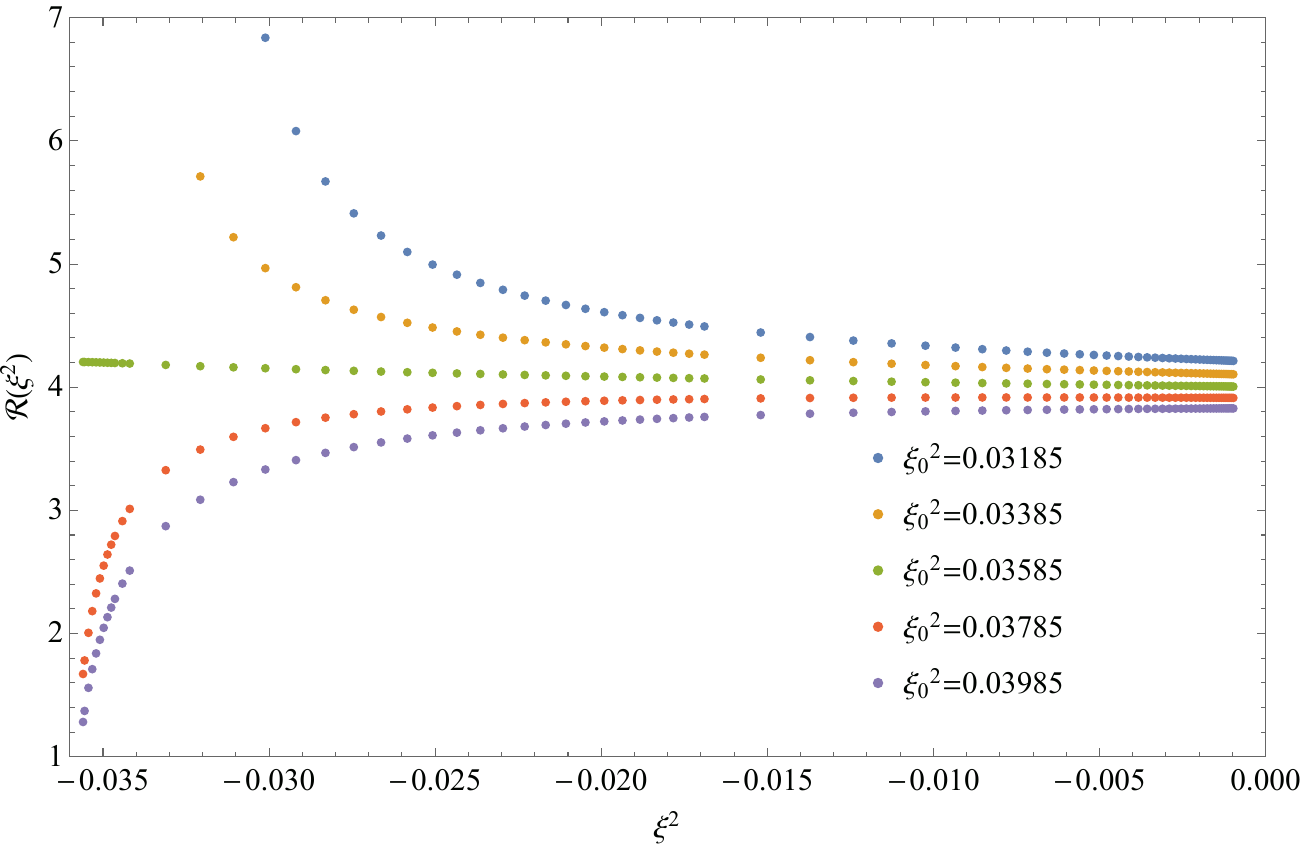}
\end{minipage}
\begin{minipage}[htb]{0.49\textwidth}
\centering
\includegraphics[width=1.0\textwidth]{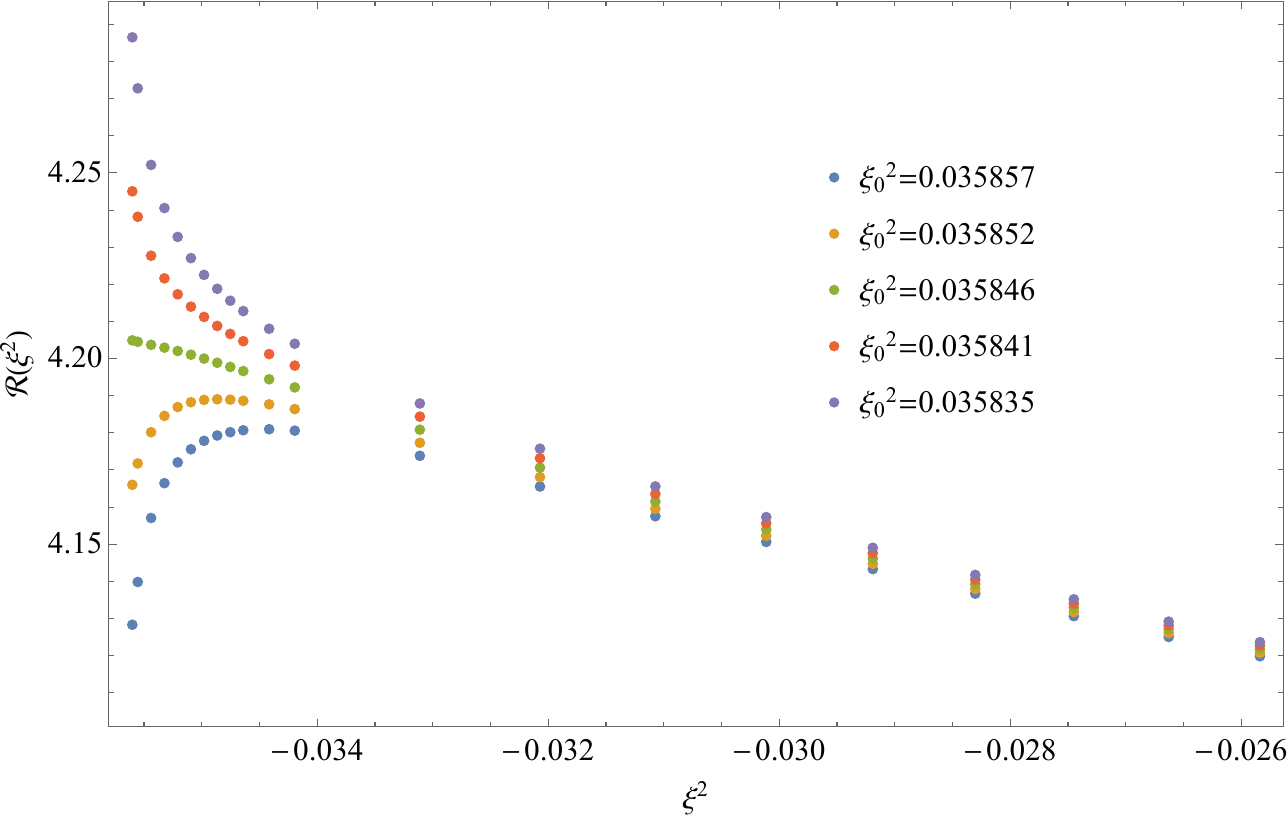}
\end{minipage}
\label{RdataYLcritical}
\end{figure}
Since the location $\xi_0^2$ is known only approximately, in Fig.\ref{RdataYLcritical} we plot the ratio $\mathcal R(\xi^2)$ computed with several values of the parameter $\xi_0^2$, close to the previous estimates. This allows to refine the location of the YL critical point,
\begin{eqnarray}\label{ylpestimate}
\xi_0^2 = 0.035846(4)
\end{eqnarray}
where the estimated error relates to low accuracy of the our data for $M(\xi^2)$ near the YL point (see previous subsection).

\subsection*{Singular Expansion}

\begin{figure}[!h]
\caption{3-parameter fitting of $M(\xi^2)$ from the singular expansion, via Eq.\eqref{Mexpansion}.}
\centering
\includegraphics[width=0.95\textwidth]{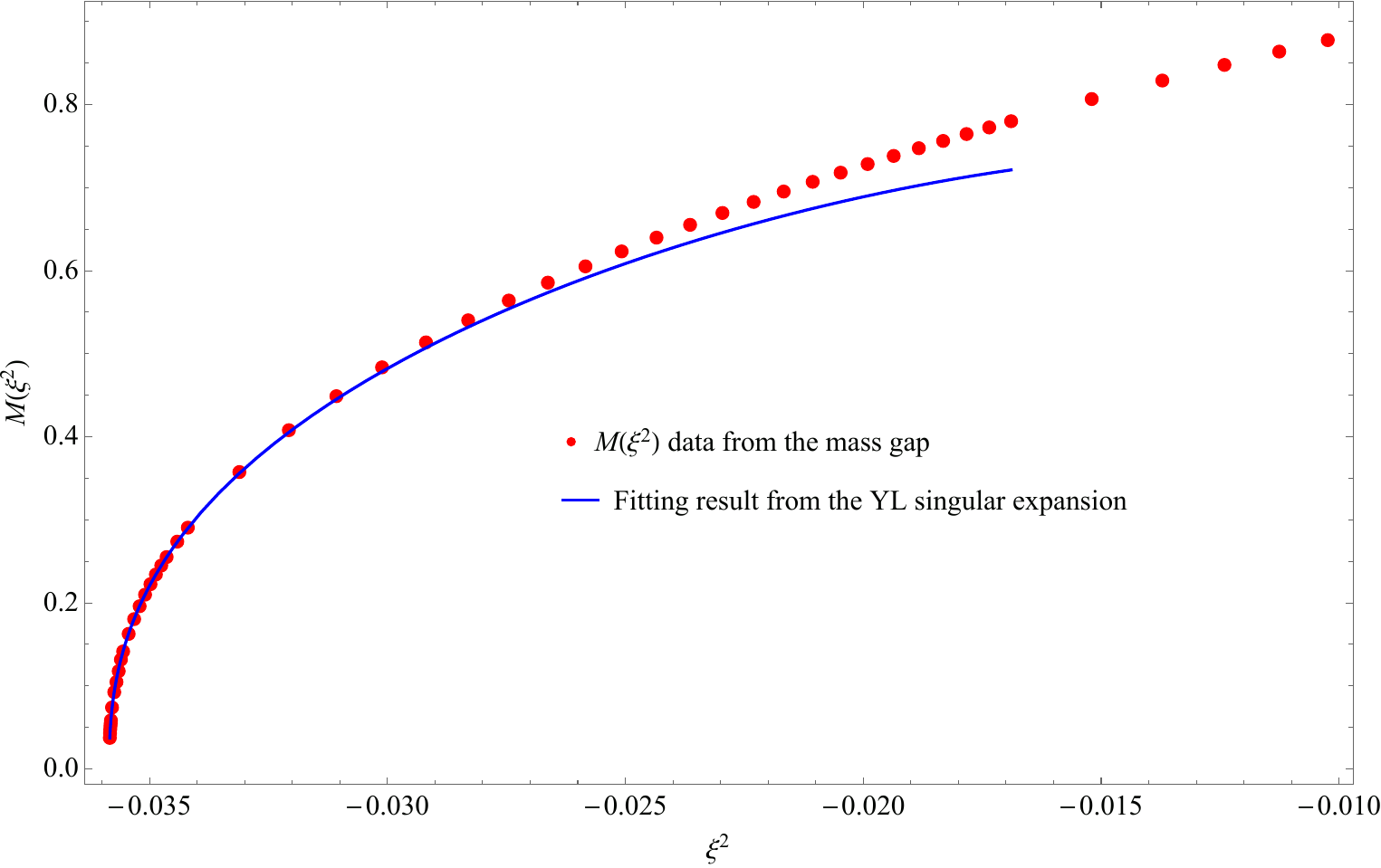}
\label{MassFittingBelowLY}
\end{figure}

The plots in Fig.\ref{RdataYLcritical} exhibit an approximate value of the leading coefficient $b_0 \approx 4.2$
\footnote{This estimate lacks good precision because the our data for $\mathcal R(\xi^2)$ have low accuracy in the close vicinity of the critical point. Better estimate, Eq.\eqref{b0b1c0} below, is based on more accurate data somewhat away from the YL point.}
in the singular expansion \eqref{Mexpansion} of $M(\xi^2)$ near the YL point, and further coefficients can be obtained by direct fitting the data to
\eqref{Mexpansion}. The best fit in the interval $0<\xi^2+\xi_0^2<0.01$
%still doesn't look right. compare to Fig.14
is obtained with
\begin{equation}\label{b0b1c0}
b_0 = 4.228 \pm 0.005\,,\quad b_1 = 21.9 \pm 0.9 \,,\quad c_0 = -14.4 \pm 0.6\,.
\end{equation}
The quality of the fit is shown in Fig.\ref{MassFittingBelowLY}.
%Should we show R instead?
With these values, the relations \eqref{Mexpcoeffs} allow one to estimate some coefficients in the expansions \eqref{lambdaexp}, \eqref{alphaxi},
\begin{eqnarray}\label{lambdaestimate}
\lambda_1 = 3.089 \pm 0.008 \,, \qquad \lambda_2 = 38.4 \pm 1.6
\end{eqnarray}
and
\begin{eqnarray}
\alpha_0 = - 1.32 \pm 0.05\,.
\end{eqnarray}
The last number agrees with the estimate \eqref{alphaestimate}, providing another independent cross-check of the latter. Also, the estimate
of these values are in agreement with the result of Ref.\cite{fonseca2003ising} (see Eq.(7.3) there).

In principle, the singular expansion \eqref{Mexpansion} can be continued further,
\begin{eqnarray}\label{MassSingularExpansionExtensionFormula}
&&M(\xi^2) = (\xi^2+\xi_0^2)^{5/12}\,\left[b_0 + b_1\,(\xi^2+\xi_0^2) + b_2\,(\xi^2+\xi_0^2)^2 + ...\right] + \\
&&\,\qquad\qquad (\xi^2+\xi_0^2)^{5/4}\,\left[c_0 + c_1\,(\xi^2+\xi_0^2) + ...\right] + (\xi^2+\xi_0^2)^{11/4}\,\left[d_0 +  ...\right] + ...\nonumber
\end{eqnarray}
where the last of the exposed terms represents the contribution of the operator $\Xi$ in the effective action \eqref{aeff1} (and
$d_0 \propto \beta_0$), while the contributions of yet higher operators are represented by the final dots. Unfortunately, the accuracy of our TFFSA data for $M(\xi^2)$ is hardly sufficient for a reliable estimates of the higher coefficients in this expansion.

\begin{comment}

\begin{figure}[htb]
\caption{Real and imaginary part of the spectrum at $\xi^2 = -1.853 < -\xi^2_0$.}
\centering
\begin{minipage}[htb]{0.49\textwidth}
\centering
\includegraphics[width=1.0\textwidth]{./Plots/WSSpectrumReExample.pdf}
\end{minipage}
\begin{minipage}[htb]{0.49\textwidth}
\centering
\includegraphics[width=1.0\textwidth]{./Plots/WSSpectrumImExample.pdf}
\end{minipage}
\label{WSSpectrumExample}
\end{figure}
\end{commentp}

\end{comment}

\section{Analyticity of $M(\xi^2)$ and Dispersion Relation.}\label{Sec:Mdisp}

In this Section we verify the analyticity of the function $M(\xi^2)$ at complex values of $\xi^2$. It is natural to assume that $M(\xi^2)$ is analytic on the whole complex plane of $\xi^2$ with the branching singularity at the YL point and the branch cut from $-\infty$ to $-\xi_0^2$, as shown in Fig.\ref{complexxiphasediagram}\footnote{We limit attention to the principal sheet of the Riemann surface. It is possible - and in fact likely - that the theory has other critical singularities when analytically continued under the
branch cut in Fig.\ref{complexxiphasediagram}, see \cite{Fateev:2009jf}.}. This is not a trivial assumption. Although there are no other critical points
on the principal sheet in Fig.\ref{complexxiphasediagram}, the particle masses can have algebraic singularities similar to the "level crossings"
in Schroedinger equation with complex parameters. (In fact, the higher masses $M_n(\xi^2), n=2,3,...$ do have such singularities,
as we will discuss elsewhare.) So, our assumption here is that the lightest mass $M(\xi^2)$ (and hence the correlation length)
has no singularities other than the YL point. In addition, $M(\xi^2)$ enjoys the asymptotic behavior
\begin{eqnarray}\label{Mass0}
M(\xi^2) \ \to \ M^{(0)}\,[\xi^2]^\frac{4}{15}
\quad \text{when} \quad \xi^2 \ \to \ +\infty
\end{eqnarray}
which follows from the fact that at non-zero $h$ the mass $M$ has finite limit $m_0\,|h|^{8/15}$ at $m=0$ (and in fact analytic at this point), with the coefficient $M^{(0)}$ known from the integrable IFT at $m=0$ (analytic expression is presented in Appendix \ref{AppendixNumbers},
Eq.\eqref{M0exact}.). With this analyticity assumptions, the function $M(u)$ (below in this section we use the notation $u:=\xi^2$)
must obey the dispersion relation
\begin{eqnarray}\label{MassDisperionRelationxi2plane}
M (u) = 1 + u\,\int_{\xi_0^2}^{+\infty} \frac{dv}{\pi} \frac{ \Im m \, M (-v + i0)}{v(v + u)}\,,
\end{eqnarray}
which expresses its values at all complex $u$ in terms of the discontinuity
\begin{eqnarray}
\text{Disc}\, M (-v) =  {M} (-v + i0) - {M} (-v - i0) = 2 i\, \Im m \, {M} (-v + i0).
\end{eqnarray}
across the branch cut in Fig.\ref{complexxiphasediagram}.

To verify the dispersion relation we need to build some approximation for the discontinuity. In principle, the imaginary part
of $M(-u)$ can be determined by direct analysis of TFFSA numerics for four lowest levels $E_n(R)$ at $\xi^2 < -\xi_0^2$.
In this domain of $\xi^2$ all energies $E_n(R)$ become complex at sufficiently large $R$, and form complex conjugated pairs.
The two "lowest" (in the sense of the real parts of $E_n(R)$) levels at large $R$ approach the linear asymptotic forms
\begin{eqnarray}\label{twovacua}
E_0(R) \to F_{+}R\,, \qquad E_{1}(R) \to F_{-}R
\end{eqnarray}
exponentially fast, with the slopes complex conjugate to each other, $F_{-}=F_{+}^*$. In the limit $R=\infty$ these levels represent two "degenerate" (again, in the sense of the real parts of $E$) vacua, the manifestation of the spontaneous breakdown of certain discrete symmetry. The next two levels exponentially approach the asymptotic forms
\begin{eqnarray}\label{twovacuaparticles}
E_2(R) \to F_{+}R +M_{+}\,, \qquad E_3(R)\to F_{-}R + M_{-}\,,
\end{eqnarray}
where again $M_{+}$ and $M_{-}$ are complex conjugate to each other. The corresponding states may be interpreted as the one-particle excitations over the vacua \eqref{twovacua}, with $M_{+}$ and $M_{-}$ interpreted as the associated complex masses. The functions
$M_\pm(u)$ defined this way give the values $M(u\pm i0)$ of the analytic continuation of $M(u)$ at the upper and lower edges of the
branch cut in Fig.\ref{complexxiphasediagram}. When $u$ is taken sufficiently far away from  $-\xi_0^2$ asymptotic behavior \eqref{twovacua},\eqref{twovacuaparticles} are well visible within the domain where TFFSA returns accurate data, and $\Im m \, M(u)$ can be estimated from these data. However, the accuracy of this estimate is not particularly good, especially when we get closer to the
critical point. Therefore, we employed another approach to estimate $\Im m \, M(u)$.

Our approximation is based on two complimentary sets of data. One is the singular expansion \eqref{Mexpansion}, which allows one to construct an approximation for the the discontinuity in some domain close to the YL critical point,
\begin{eqnarray}\label{ImMexpansion}
\Im m \, M(-u+i0) = (u-\xi_0^2)^\frac{5}{12} \big[ \, b_0' + b_1'\,(u-\xi_0^2) + c_0'\,(u-\xi_0^2)^\frac{5}{6} + \text{higher terms} \big]
\end{eqnarray}
where $b_0' = b_0\,\sin(5\pi/12)$, $b_1' = b_1\,\sin(17\pi/12)$, $c_0' = c_0\,\sin(5\pi/4)$, etc. On the other hand, $M(\xi^2)$
admits another expansion, convergent at large $\xi^2$, of which \eqref{Mass0} represents just the leading term. If measured in the
units of $|h|^{8/15}$, the mass $M$ admits Taylor expansion
\begin{eqnarray}\label{Mseries}
M/|h|^{8/15} = M(\xi^2)/(\xi^2)^{4/15} = M^{(0)}\, + M^{(1)}\,\eta + M^{(2)}\,\eta^2 +M^{(3)}\,\eta^3 + ...
\end{eqnarray}
in powers of the variable\footnote{The scaling parameter $\eta$ allows to chart the analytic picture uniting both High-T and Low-T regimes. See \cite{fonseca2003ising} for details.}
\begin{eqnarray}\label{etaxi}
\eta: = -(\xi^2)^{-4/15} = -m/|h|^{8/15}\,.
\end{eqnarray}
In principle, the coefficients $M^{(n)}$ can be computed via the perturbation theory around the integrable theory \eqref{ift} with
$m=0$. Thus, the first two coefficients $M^{(0)}$ and $M^{(1)}$ are known exactly (we present the closed form expressions in Appendix \ref{AppendixNumbers}). The higher $M^{(n)}$ was never computed exactly, and we determined few further coefficients numerically, using TFFSA data at small $\eta$. Thus we have
\begin{eqnarray}\label{MnCoeffs}
M^{(0)}=4.404908...\,, \quad M^{(1)}=1.29531...\,, \quad M^{(2)}=0.2002\,, \quad (M^{(3)} = -0.051)
\end{eqnarray}
At real positive $\xi^2$ the relation \eqref{etaxi} is understood in a straightforward way (the principal branch of the
power function is taken). However, it allows for analytic continuation to real negative $\xi^2$, where the variable \eqref{etaxi}
takes complex values along the rays
\begin{eqnarray}
\eta = -y\, e^{\pm i \frac{4\pi}{15}}\,,
\end{eqnarray}
with real positive $y$. The segments of these rays $y\in [0,Y_0]$, where
\begin{eqnarray}
Y_0 = (\xi_0^2)^{-4/15} = 2.4293...\,,
\end{eqnarray}
represent the images of the upper and lower edges of the branch cut in Fig.\ref{complexxiphasediagram}, respectively.
%or is it the other way round?
Therefore, the imaginary part of $M$ at the upper edge of the branch cut in Fig.\ref{complexxiphasediagram} is given by the series
\begin{eqnarray}\label{ImMseries}
\Im m\, M(-u) = M^{(0)}_\text{Im}\,y^{-1} + M^{(1)}_\text{Im} + M^{(2)}_\text{Im}\,y + M^{(3)}_\text{Im}\,y^2 + ...
\end{eqnarray}
where again $y = (-u)^{-4/15}$, and
\begin{eqnarray}
M^{(n)}_\text{Im} = M^{(n)}\, \sin\Big[ \frac{4\pi (1-n)}{15}\Big]\,.
\end{eqnarray}
The series in the r.h.s. of \eqref{ImMseries} is expected to converge at $y<Y_0$, and at $y>Y_0$ the imaginary part turns to zero.

\begin{figure}[!h]
\caption{Plots of $\Im m\, \M(y)$ above the upper edge of YL branch cut, $0<y<Y_0$. The red, blue and purple solid lines correspond to the analytical continuation via Eq.\eqref{ImMexpansion}, Eq.\eqref{ImMseries} and Eq.\eqref{MassSingularContinuationFormulaExtended}. The red bullets are given by fitting the imaginary part of $\Delta E(R)$ for $\xi^2 < -\xi_0^2$.}
\centering
\includegraphics[width=0.75\textwidth]{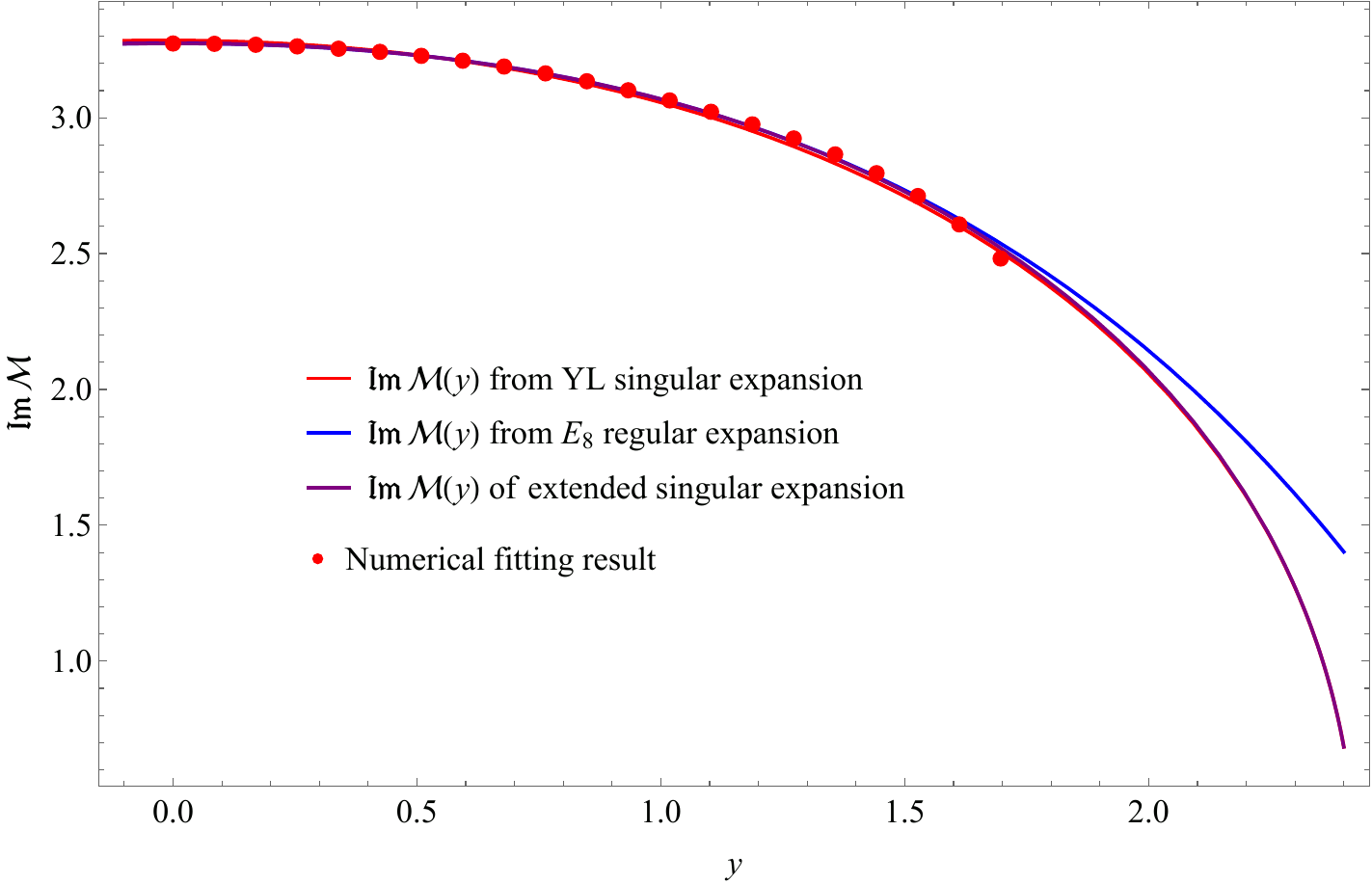}
\label{ImMmatching}
\end{figure}

We found it convenient to evaluate the dispersion integral in \eqref{MassDisperionRelationxi2plane} using the above
variable $y: = (-u)^{-4/15}$ instead of $u$ (one of the advantages being that in this variable the integral extends over the finite domain $[0:Y_0]$). The dispersion relation takes the form
\begin{equation}\label{MassDisperionRelation}
M(u) = \Big( 1 + \frac{15}{4\pi} \int_0^{Y_0} \frac{y^{7/4} \Im m \, \M (-y+i0)}{y^{15/4} + u^{-1}} dy \Big)\,,
\end{equation}
where $\Im m \M(y)$ denotes $y \Im m M$ expressed through the variable $y$.
Our approximation for $\M(y)$ is
based on the singular expansion
\begin{eqnarray}\label{MassSingularContinuationFormulaExtended}
\Im m \M(y)&=& (Y_0-y)^{\frac{5}{12}}\Big[ {\tilde b}_0 \sin(\frac{5\pi}{12}) + {\tilde b}_1 (Y_0-y) \sin(\frac{17\pi}{12}) + {\tilde b}_2 (Y_0-y)^2 \sin(\frac{29\pi}{12})  \nonumber\\
 & + & {\tilde c}_0\,(Y_0-y)^{\frac{5}{6}} \sin(\frac{5\pi}{4}) +   {\tilde c}_1 (Y_0-y)^{\frac{11}{6}} \sin(\frac{9\pi}{2}) +  {\tilde d}_0 (Y_0-y)^{\frac{7}{3}} \sin(\frac{11\pi}{4}) \Big]
\end{eqnarray}
for $y<Y_0$, which is equivalent of the first six terms of the expansion \eqref{ImMexpansion} (including three of the
"higher terms", see Eq.\eqref{Mexpansion}). The coefficients ${\tilde b}_n, {\tilde c}_n, {\tilde d}_n$ are related to the
coefficients $b_n, c_n, d_n$ in \eqref{MassSingularExpansionExtensionFormula} in a straightforward way, e.g.
\begin{eqnarray}\label{bctilde}
{\tilde b}_0 = \Big(\frac{15}{4}\Big)^\frac{5}{12} Y_0^{-\frac{47}{48}}\,b_0 \,, \ {\tilde c}_0= \Big(\frac{15}{4}\Big)^\frac{5}{4} Y_0^{-\frac{79}{16}}\,c_0 \,,\
{\tilde b}_1 = \Big(\frac{5}{4}\Big)^\frac{5}{12} \left(\frac{360 b_1 + b_0 Y_0^{15/4}}{32 \cdot 3^{\frac{7}{12}} Y_0^{275/48}}\right) \,.
\end{eqnarray}
We use the previous estimates \eqref{b0b1c0} to fix the coefficients ${\tilde b}_0$, ${\tilde b}_1$, and ${\tilde c}_0$
according to \eqref{bctilde}, and then adjust the remaining coefficients in \eqref{MassSingularContinuationFormulaExtended}
to match the first three terms of the expansion \eqref{ImMseries} around $y=0$. The resulting numerical values are displayed in Table.\ref
{6ParametersSingularCoefficients}\footnote{We would like to stress that the values of ${\tilde b}_2, {\tilde c}_1, {\tilde d}_0$
in Table \ref{6ParametersSingularCoefficients} are not to be regarded as meaningful estimates of actual higher order coefficients in the expansion \eqref{Mexpansion}, and thus estimates of yet higher irrelevant couplings in \eqref{aeff0}. Rather, they are just elements of our approximation \eqref{MassSingularContinuationFormulaExtended} designed to match the data \eqref{MnCoeffs}.}. The plots in Fig.\ref{ImMmatching} show how this approximation compares with direct numerical estimates of $\Im m \M(-y+i0)$.
\begin{table}[htp]
\begin{center}
\begin{tabular}{|c|c|c|c|c|c|}
\hline
${\tilde b}_0$ & ${\tilde b}_1$ & ${\tilde b}_2$ & ${\tilde c}_0$ & ${\tilde c}_1$ & ${\tilde d}_0$ \\
\hline
$3.0754$ & $0.8932$ & $-0.9767$ & $-0.9412$ & $1.026$ & $0.3329$ \\
\hline
\end{tabular}
\end{center}
\caption{Numerical values of the 6-parameter singular expansion coefficients, see Eq.\eqref{MassSingularContinuationFormulaExtended}.}
\label{6ParametersSingularCoefficients}
\end{table}

With this approximation for the discontinuity, the dispersion relation \eqref{MassDisperionRelation} gives (approximate)
values of $M(\xi^2)$ at all complex $\xi^2$ in Fig.\ref{complexxiphasediagram}. For real $\xi^2$ (both positive and negative) the result of direct
numerical integration is shown in Fig.\ref{MassFittingFullxi2}.

Some quality checks of this approximation are easy to perform. Consider the expression
\begin{eqnarray}\label{MDispY0}
\Delta M: = 1 -\xi^2_0 \int_{\xi_0^2}^{+\infty} \frac{du}{\pi} \frac{ \Im m \, M (-u + i0)}{u(u -\xi^2_0)}
\end{eqnarray}
which represents, according to \eqref{MassDisperionRelationxi2plane}, the value $M(-\xi_0^2)=0$. That is, with exact $\Im m {M}(-u+i0)$ this expression must return zero. With our approximation, the numerical integration in \eqref{MDispY0} yields $\Delta M =0.00251378$. Baring in mind that the integrand in \eqref{MDispY0} takes values $\sim 1$ in the integration domain, this is reasonably small error.

Note that while the specific values of the coefficients $M^{(0)}$ and $M^{(2)}$ in the expansion \eqref{Mseries} (Eq.\eqref{MnCoeffs}) are incorporated, through the coefficients $M^{(0)}_{\text{Im}}$ and $M^{(2)}_{\text{Im}}$, into our approximation, the coefficient $M^{(1)}$ does
not contribute to the expansion \eqref{ImMseries} at all ($M^{(1)}_{\text{Im}}$ vanishes no matter what $M^{(1)}$ is). However, given the
discontinuity $2i\,\Im m {\cal M}(y)$, it is possible to recover the coefficient $M^{(1)}$ through the dispersion relation \eqref{MassDisperionRelation}. It is not difficult to derive the relation
\begin{eqnarray}\label{M1Disp}
M^{(1)} = -1 + \frac{15\,M^{(0)}_\text{Im}}{4\pi\,Y_0} - \frac{15}{4\pi}\,\int_{0}^{Y_0}\,\frac{\Im m{\cal M}(-y)-M^{(0)}_\text{Im}}
{y^2}\,dy\,.
\end{eqnarray}
With exact $\Im m {\cal M}(y)$ this expression must return the exact $M^{(1)}$ given in Eq.\eqref{MnCoeffs}. Numerical evaluation of the integral in \eqref{M1Disp} with our approximation results in the number
\begin{equation}
M^{(1)} = 1.29591 \,,
\end{equation}
reasonably close to the exact value \eqref{M1exact}.

Finally, the function $M(\xi^2)$ is regular at $\xi^2=0$, and admits power series expansion
\begin{eqnarray}\label{Mxiexpansion}
M(\xi^2) = -m + \mu_2 \,\xi^2 + \mu_4\,\xi^4 + ...
\end{eqnarray}
convergent in some domain around $\xi^2=0$. The coefficient $\mu_2$ is known exactly, through the perturbation theory of IFT around the point $\xi^2=0$ (see Ref.\cite{fonseca2003ward}), while $\mu_4$ can be estimated by fitting the TFFSA data near at small $\xi^2$. On the other hand, it is straightforward to derive the identities
\begin{equation}\label{MassDisperionRelationCoefficients}
\mu_{n} = (-)^{n+1}\int_0^{Y_0} y^{\frac{15n-8}{4}} \Im m \, \M(-y) \, dy\,.
\end{equation}
Then the numerical integrations yields
\begin{eqnarray*}
\mu_2=10.7485\,, \qquad \mu_4=-96.9807\,,  \qquad \mu_6=1455.36\,,
\end{eqnarray*}
to be compared with the exact value of $\mu_2$, Eq.\eqref{B3}, and the estimates $\mu_4 = 97.22$ and $\mu_6 = 1396$ obtained by direct fitting of $M(\xi^2)$ at small $\xi^2$.

\begin{figure}[!h]
\caption{Summary of the plots of ${M}(\xi^2)$ from measurement and dispersion relation.}
\centering
\includegraphics[width=1.0\textwidth]{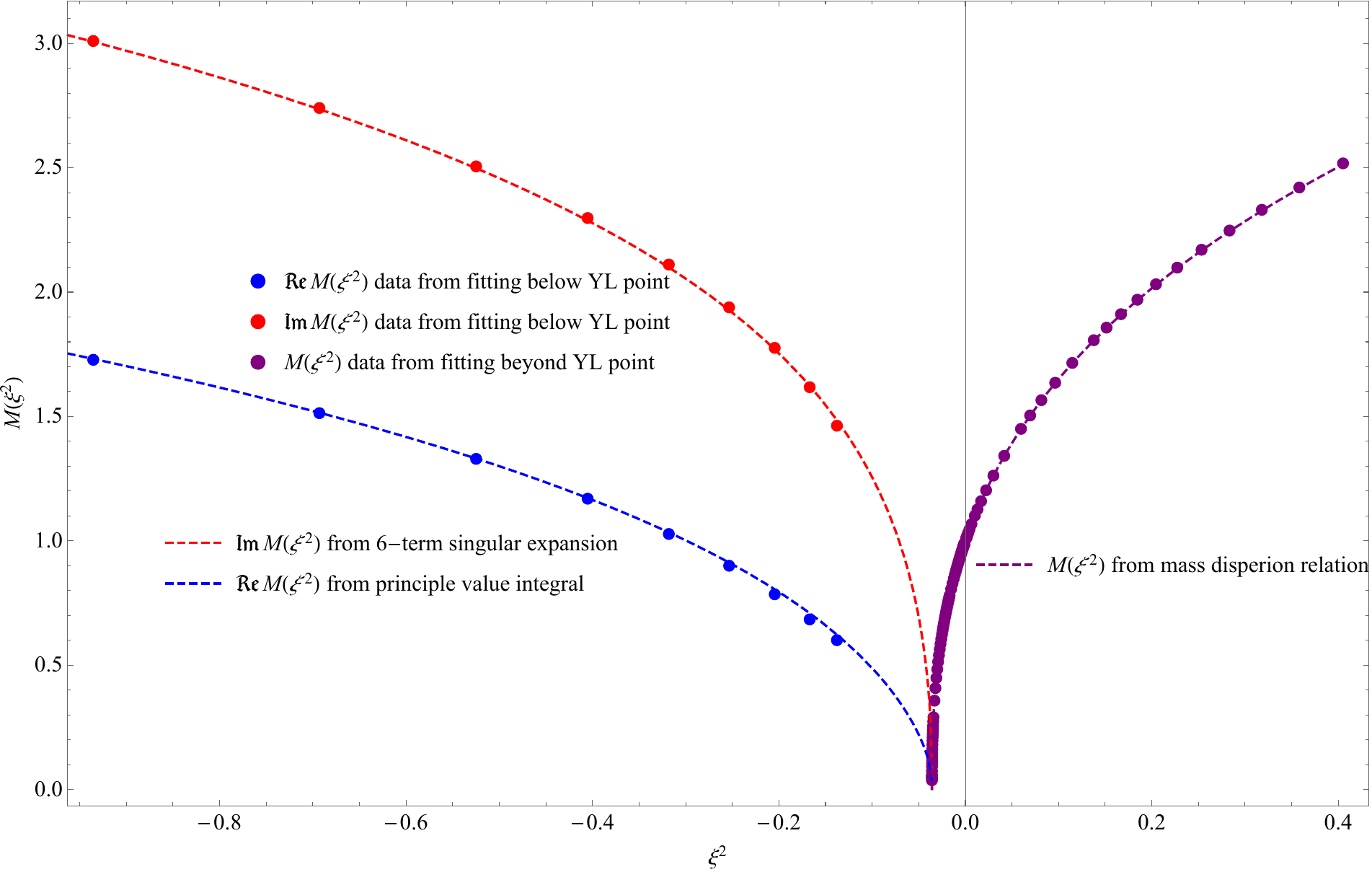}
\label{MassDispersionAllxi2}
\end{figure}

The plots in Fig.\ref{MassDispersionAllxi2} show combined data from the gap fitting and the results of the numerical evaluation
of the integral in \eqref{MassDisperionRelation}. These results, as well as the above consistency checks, strongly support our conjecture about relatively simple analytic structure of the mass $M(\xi^2)$ at complex $\xi^2$ shown in Fig.\ref{complexxiphasediagram}.

\section{Summary and Discussion}\label{conclusion}
In this work we continued the study of IFT, Eq.\eqref{ift}, in pure imaginary magnetic field $h$, with particular emphasis on the
effective action designed to describe the close vicinity of the Yang-Lee critical point $h/|m|^{15/8}=\pm i\xi_0$. The effective action describes the RG flow close to the "massless flow" from the Ising fixed point down to the Yang-Lee fixed point, see Fig.\ref{RGflowsPicture}. It has the form \eqref{aeff0}, which is the Yang-Lee QFT \eqref{ylqft} deformed by an infinite tower of irrelevant operators. Of those the most important are the lowest descendants of $I$ and $\phi$, the operators $T{\bar T}$, and $\Xi$, as exhibited in the "truncated" effective action \eqref{aeff1}. The couplings $\lambda$, $\alpha$ and $\beta$ all depend on the scaling parameter $\xi^2$ in a nontrivial way, but admit the power series expansions \eqref{lambdaexp},\eqref{alphaxi},\eqref{betaxi}. We use numerical data for few lowest finite-size energy levels (obtained via TFFSA) to estimate some leading coefficients of these expansions. We also give a refined numerical estimate of the position of the Yang-Lee critical point, Eq.\eqref{ylpestimate}.

Our estimate of $\lambda_1$ and $\alpha_0$ in Eq.\eqref{lambdaestimate} and \eqref{alphaestimate} are in agreement with the previous estimates in \cite{fonseca2003ising}, but we believe have better precision. The estimate \eqref{betaestimate} of $\beta_0$ is new.
We believe that the estimate \eqref{ylpestimate} of $\xi_0^2$ obtained here is more accurate than that given in \cite{fonseca2003ising}.

The enhanced precision in due to a number of technical improvements. We use TFFSA data obtained at the truncation level $L=13$,
whereas \cite{fonseca2003ising} uses levels up to $12$. We employ the $T{\bar T}$-deformation formula \eqref{ealpha} in order to take into account corrections of higher order in the coupling $\alpha$ in \eqref{aeff1}. In addition, the integrability of the Yang-Lee QFT \eqref{ylqft} and the Thermodynamic Bethe Aanzats was used to construct the form \eqref{E1-TBA} for fitting the finite size energy levels in close vicinity of the YL critical point.

Large part of our analysis relies on numerical solution (via TFFSA) of the IFT in its continuous version \eqref{ift}. Alternative approach to the Ising universality class based on numerical solution of the lattice Ising Model in a magnetic field, through
the corner transfer matrix technique, was developed in \cite{Mangazeev:2008wg,Mangazeev:2010ye}. Very recently that approach was extended to the case of pure imaginary magnetic field in \cite{BazhanovYL}, where in particular the position of the YL singularity was estimated as $\xi_0^2 \approx 0.035868$, which deviates from our estimate \eqref{ylpestimate} only in the last two digits.

At generic values of parameters $m$ and $h$ the Ising Field Theory \eqref{ift} is not integrable. This statement almost certainly applies to all real $\xi^2$ except the points $\xi^2=0$ and $\xi^2=\infty$, where \eqref{ift} reduces to special Integrable QFT's. The non-integrability can be seen e.g. in the presence of inelastic scattering processes explicitly exhibited e.g. in \cite{Zamolodchikov:2011wd},\cite{Gabai:2019ryw} at small $\xi^2$. As the YL QFT \eqref{ylqft} (as well as its $T{\bar T}$ deformation) is integrable, non-integrability of IFT at $\xi^2$ close to $-\xi_0^2$ implies the presence of integrability breaking operators among the tower of irrelevant operators in the effective action \eqref{aeff0}. The significance of the operator $\Xi$ in \eqref{aeff1} is that it is the lowest dimension operator breaking the integrability. We plan to say more on non-integrable features of IFT near YL criticality in the future work \cite{XuEtAl2022}.

Our estimates of the parameters in \eqref{aeff1} was based on the analysis of the lowest finite-size levels $E_0(R), E_1(R), E_2(R)$
which behave in relatively simple manner, see Figs.\ref{SpectrumCritical} and \ref{SpectrumExample}. The levels $E_3(R)$ and $E_4(R)$ shown in Fig.\ref{SpectrumCritical}, exhibit more intricate behavior, with two "level crossings" where these eigenvalues collide and turn into the complex-conjugate pair. We show that the level crossing at greater $R\approx 13$ is nicely explained
as the integrability-breaking effect of the operator $\Xi$ in the effective action \eqref{aeff1}. The numerical match shown in
Fig.\ref{Fig.C34c} confirms our estimates of the coupling parameters. Yet higher levels $E_n(R), n=5,6,...$ (not presented in Fig.\ref{SpectrumCritical}) generally show even more complicated behavior, forming a web of real and complex-conjugate eigenvalues with many level crossings. Understanding of of this behavior in terms of the effective action \eqref{aeff0} remains an interesting open problem.

Our numerical data for the mass $M(\xi^2)$ allowed us to confirm the simple analyticity conjecture of this function at complex
$\xi^2$. Specifically, we verified the dispersion relation \eqref{MassDisperionRelationxi2plane} which expresses this analyticity.

Let us make a remark on higher-dimension irrelevant operators not included in \eqref{aeff1}. The higher level descendants
of $I$ potentially appearing in \eqref{aeff0} are $X_5 = L_{-2}^3 {\bar L}_{-2}^3 I$, $X_7 = L_{-2}^4 {\bar L}_{-2}^4 I$
and $X_7 = L_{-2}^4 {\bar L}_{-2}^4 I$, which fill the slots $l=6, 8$ and $10$ in Table \ref{Tab:dimensions}. These are all representatives of an infinite series of operators $X_s$, $s=1,5,7,11,...$ (odd integers not divisible by 3) introduced in Ref.\cite{smirnov2017space}. These operators generate an infinite-dimensional "generalized $T{\bar T}$ deformations" which preserve integrability. Although for the generalized deformations there is no formula describing the dependence of the finite-size energies on the deformation parameters $\alpha_s$ as simple and efficient as \eqref{ealpha}, the special properties of the operators $X_s$ give at least some control over the effect of these operators in the effective action \eqref{aeff0}. In particular, the corresponding deformation of the S-matrix is known, and can be used to take account for the contributions of the operators $X_s$ through the TBA technique.
Including the contributions of these operators may significantly improve the power of the effective action \eqref{aeff0}, especially for the higher energy levels. We hope to return to this question in the future.

\section*{Acknowledgements}

AZ acknowledges discussions with F.Smirnov, and HLX thanks helpful discussion with R.Shrock. We thank V.Bazhanov for sharing some results of his work \cite{BazhanovYL} prior to publication. Research of AZ is partly supported by NSF under grant PHY-191509.

\section*{Appendix}

\appendix
\section{Matrix elements of descendent operators at criticality.}\label{AppendixCPT}

Here we derive the diagonal matrix elements
\begin{eqnarray}\label{Xinndef}
\Xi_{nn}:=\langle n | \Xi(0) | n \rangle
\end{eqnarray}
(we set $R=2\pi$ here) in Yang-Lee CFT $\mathcal{M}_{2/5}$ for the lowest four levels $n$, quoted in \eqref{Xinn}. The field $\Xi$ is the descendant of $\phi$ defined in \eqref{Xidef}. The terms in \eqref{Xidef} involving $L_{-1}$ and/or ${\bar L}_{-1}$ bring zero contributions to the diagonal matrix elements \eqref{Xinndef}, and here we set simply $\Xi=L_{-4}{\bar L}_{-4}\phi$.

As usual, for a primary field $\mO_P$ its descendant $L_{-n}\mO_P$ is defined as the integral
\begin{eqnarray}
L_{-n}\mO_P(z_0,\zb_0) = \oint_{\mathcal C_{z_0}} \frac{dz}{2\pi i}\,(z-z_0)^{1-n}\,T(z) \,\mO_P(z_0,\zb_0)
\end{eqnarray}
where $z$ is a local complex coordinate covering some neighborhood of the point $z_0$, and integration goes over a small contour
encircling this point (and similar expression exists for ${\bar L}_{-n}\mO_P$). On the r.h.s. one can replace $(z-z_0)^{1-n}$ in the integrand by any function $U_n(z-z_0)$ having the Laurent expansion
\begin{eqnarray}\label{ULaurent}
U_n(z) = \frac{1}{z^{n-1}} + O(z^3)
\end{eqnarray}
without altering the result. This is because the terms $(z-z_0)^{1+n}$ generate contributions $L_{n}\mO_P$, which for primary
$\mO_P$ all vanish for $n>1$.

Now let $z=\tx+i\ty$ be the global complex coordinate on a cylinder in Fig.\ref{cylinder}, of circumference $R=2\pi$. It is possible to construct functions $U_n(z)$ which are $2\pi$ periodic, $U(z+2\pi)=U(z)$, analytic on the cylinder everywhere except the point $0$, where
they have the Laurent expansion \eqref{ULaurent} (these conditions of course do not fix the functions uniquely, but any
choice would do). We only need the function $U_4(z)$, which can be chosen in the form
\begin{eqnarray}\label{U4}
U_4(z)=\frac{1}{8} \,\frac{\cos(z/2)}{\sin^3(z/2)}+ \frac{1}{240}\,\sin(z)
\end{eqnarray}
Note that it admits two different Fourier expansions
\begin{eqnarray}\label{U4plus}
&&U_4(z)-\frac{1}{240}\,\sin(z) = \ \ \frac{1}{2i}\,\big( e^{-iz} + 4\,e^{-2iz}+9\,e^{-3iz} + 8\,e^{-4iz} + \dots \big)\qquad \Im m \, z>0\,,\\
&&U_4(z)-\frac{1}{240}\,\sin(z) = -\frac{1}{2i}\,\big(e^{+iz} + 4\,e^{+2iz}+9\,e^{+3iz} + 8\,e^{+4iz} + \dots \big)\qquad \Im m \, z<0\,,\label{U4minus}
\end{eqnarray}
convergent in the upper and lower half-cylinders, respectively.

Consider a matrix element
\begin{eqnarray}\label{fL4i}
\langle f | L_{-4}\phi(0) | i \rangle = \oint_{\mathcal C_0} \frac{dz}{2\pi i}\,U_4(z)\,\langle f | T(z) \phi(0,0)
| i \rangle
\end{eqnarray}
between two states $| i \rangle$ and $| f \rangle$ from the space of states of the CFT on the cylinder; here $U_4(z)$ is the function \eqref{U4}. The contour $\mathcal{C}_{0}$ can be deformed into the combination of two contours, $\mathcal{C}_{-}$ and $\mathcal{C}_{+}$, where $\mathcal{C}_{-}$ goes around the cylinder in Fig.\ref{cylinder} below the insertion point $z=0$, while $\mathcal{C}_{+}$ does the same just above the insertion point. Then, combining the expansions \eqref{U4plus} and \eqref{U4minus}
with \eqref{cylinderL}, and with some elementary algebra, \eqref{fL4i} is transformed to
\begin{eqnarray}\nonumber
&&\langle f | L_{-4}\phi(0) | i \rangle = -\frac{\Delta_\phi}{480}\,\langle f | \phi(0)| i \rangle + \qquad\\
&&\qquad \frac{1}{2}\,\langle f | \phi(0) \Big( \bL_{-1} + 4  \bL_{-2} +{9}  \bL_{-3} + 16  \bL_{-4} + \cdots \Big) |i \rangle + \\
&&\qquad\frac{1}{2}\,\langle f | \Big( \bL_{+1} + 4  \bL_{+2} +{9}  \bL_{+3} + 16  \bL_{+4} + \cdots \Big)\phi(0) |i \rangle\,.
\nonumber
\end{eqnarray}
Here dots represent terms involving $\bL_{n}$ with $n<-4$ and with $n>4$.

For the matrix element of $\Xi = L_{-4}{\bar L}_{-4}\phi$ the operator ${\bar L}_{-4}$ can be handled in the same manner. For the
diagonal matrix elements between the states \eqref{CFTstates} a little more algebra yields
\begin{gather}
\langle  0 | \Xi(0) |  0 \rangle=\frac{\mathbb{C}_{\phi\phi}^{\phi} }{1200^2} \approx 1.327 \times 10^{-6}\,,\\
\langle  2 | \Xi(0) |  2
\rangle=\Big(\frac{5}{2}\Big)^2\Big( \frac{601}{7500}\Big)^2\mathbb{C}_{\phi\phi}^{\phi}
=\frac{361201}{9000000}\mathbb{C}_{\phi\phi}^{\phi} \approx 0.07671 \,, \\
\langle 3 | \Xi(0) | 3 \rangle = \Big(\frac{25}{12}\Big)^2\Big( \frac{56417}{62500}\Big)^2\mathbb{C}_{\phi\phi}^{\phi} = \frac{3182877889}{900000000}\mathbb{C}_{\phi\phi}^{\phi} \approx 6.7594 \,.
\end{gather}
The off-diagonal matrix element $\Xi_{34}$ can be obtained by similar calculation, which gives:
\begin{equation}
\langle 3 |\Xi| 4 \rangle = \Big(\frac{25}{12}\Big)\Big(\frac{5}{11}\Big)\Big(\frac{19807}{12500}\Big)^2 i  = \frac{392317249}{165000000}i \approx 2.37768 i\,.
\end{equation}

\section{Some exact numbers}\label{AppendixNumbers}

Some coefficients appearing in different expansions of the function $M(\xi^2)$ are known exactly, from the perturbation theory of
\eqref{ift} around integrable points in the parameter space. Parts of these exact results are spread in different literature sources, and we present expressions obtained by combining these results (and giving them more compact form).

The expansion \eqref{Mseries} (which represents expansion of $M(\xi^2)$ at large $\xi^2\to+\infty$) can be obtained by perturbation theory around the integrable theory \eqref{ift} with $m=0$ and non-zero $h$. The first two coefficients are known in a closed form,
\begin{eqnarray}\label{M0exact}
M^{(0)} =\frac{4\pi}{\Gamma(2/3)\Gamma(4/5)\Gamma(8/15)}\,
\left[\frac{4\pi^2\,\Gamma^2(13/16)\Gamma(3/4)}{\Gamma^2(3/16)\Gamma(1/4)}\right]^{4/15}=4.404908579981566...
\end{eqnarray}
\begin{eqnarray}
&&M^{(1)} = \frac{256}{225}\,\frac{2^{1/4}}{\sqrt{3}}\,
\frac{\Gamma(1/8)\Gamma(2/5)\Gamma(4/15)\Gamma(7/15)\Gamma^2(3/4)\Gamma^2(13/16)}
{\Gamma(7/8)\Gamma(3/5)\Gamma(4/5)\Gamma(11/15)\Gamma^2(3/16)\Gamma^3(2/3)}\, \frac{\sin(4\pi/15)}{\sin(\pi/15)}\,\times \nonumber \\
&&\qquad\qquad\,\bigg[\cos\bigg(\frac{2\pi}{15}\bigg)\cos\bigg(\frac{\pi}{30}\bigg) -
\cos\bigg(\frac{\pi}{5}\bigg)\cos\bigg(\frac{7\pi}{30}\bigg)\bigg] = 1.295047691998804...\label{M1exact}
\end{eqnarray}
The form \eqref{M0exact} can be extracted from the results in \cite{Fateev:1993av}. The coefficient
\eqref{M1exact} is combined from exact results for the vacuum expectation value $\langle \varepsilon \rangle_{m=0}$ given in \cite{Fateev:1997yg} in integral form, and brought to nice closed form in \cite{Mangazeev:2008wg}, and exact results foe the form factors in \cite{Alekseev:2011my}, which we transformed to the relatively compact form above.

The coefficient $\mu_2$ in \eqref{Mxiexpansion} is known exactly from the perturbations around the integrable theory \eqref{ift} with
$h=0$ and $m<0$. Although there is no closed form in terms of conventional transcendents like \eqref{M0exact},\eqref{M1exact} above,
it can be expressed as an integral involving special solution of the Painleve III equation, see \cite{fonseca2003ward}, which allow to compute it numerically, with arbitrary accuracy,
\begin{eqnarray}
\mu_2 = 10.7619899\dots \,. \label{B3}
\end{eqnarray}

The mass $M$ also enjoys expansion in fractional powers of $h$ valid in the vicinity of the point $h=0$ and $m>0$ in \eqref{ift}. A number of exact coefficients can be found in \cite{fonseca2003ising} and \cite{Rutkevich:2009zz}.

\bibliographystyle{unsrt}

\bibliography{isingref}

\end{document}